\documentclass[%
 reprint,
 superscriptaddress,
 amsmath,amssymb,
 aps,
prab,
]{revtex4-2}

\usepackage{graphicx}
\usepackage{dcolumn}
\usepackage{bm}
\usepackage[colorlinks,linkcolor=red,anchorcolor=green,citecolor=blue,urlcolor=magenta,filecolor=cyan]{hyperref}
\usepackage{caption}
\begin{document}

\preprint{APS/123-QED}

\title{Generation of circular field harmonics in quasi-polygonal magnet apertures using superconducting canted-cosine-theta coils}

\author{Jie Li}
\affiliation{State Key Laboratory of Nuclear Physics and Technology, and Key Laboratory of HEDP of the Ministry of Education, CAPT, Peking University, Beijing, 100871, China}

\author{Kedong Wang}
\email{wangkd@pku.edu.cn }
\author{Kun Zhu}
\email{zhukun@pku.edu.cn }
\affiliation{State Key Laboratory of Nuclear Physics and Technology, and Key Laboratory of HEDP of the Ministry of Education, CAPT, Peking University, Beijing, 100871, China}
\affiliation{Institute of Guangdong Laser Plasma Techology, Baiyun,Guangzhou,510540, China}

\date{\today}

\begin{abstract}
Superconducting magnets with non-circular apertures are important for handling unconventional beam profiles and specialized accelerator applications.
This paper presents an analytical framework for designing superconducting accelerator magnets with quasi-polygonal apertures, aimed at generating precise circular field harmonics. 
In Part 1, we explore the relationship between current distributions on quasi-polygonal formers and their corresponding magnetic field harmonics.
By employing conformal mapping techniques, we establish a connection between the design of quasi-polygonal bore magnets and traditional circular bore configurations, facilitating the simplification of complex mathematical formulations.
Part 2 applies the derived current distributions to the canted cosine theta (CCT) coil magnet concept, focusing on designing analytic winding schemes that generate single or mixed circular harmonics within quasi-polygonal apertures.
This work not only advances the design of superconducting magnets but also broadens the scope of CCT technology to accommodate more complex geometries.
\end{abstract}

\maketitle

\section{Introduction}
Custom-designed magnet apertures offer significant advantages for accommodating non-conventional beam profiles and expanding the transverse sampling area, thereby improving both spatial efficiency and beam acceptance.
Quasi-polygonal geometries, including elliptical, triangular, and square shapes, are particularly effective for optimizing beam transport, especially when they are aligned with the field region prescribed by the beam’s properties. 
The beam spot shape, typically defined by the ion source and further influenced by nonlinear forces during transport, is often approximated as quasi-polygonal. 
By matching the magnet apertures to this shape, spatial efficiency can be significantly enhanced. 
For example, cross-shaped beams in laser-plasma accelerators are well-suited for quasi-square magnet designs~\cite{cross_shape}, while ribbon beams, commonly used in ion implantation and mass analyzers, benefit from slender or elliptical apertures for improved control and focusing~\cite{RibbonBeam,ion_implanter}.

Although circular apertures dominate accelerator magnet designs, quasi-polygonal apertures remain relatively uncommon and are still in the developmental phase. 
Notable applications include the CSR External-target Experiment spectrometer at IMP, China, which employs a racetrack-shaped dipole magnet to provide a large aperture for detectors while optimizing field homogeneity at reduced cost compared to traditional circular designs~\cite{chen2024design}. 
Additionally, the National Institutes for Quantum and Radiological Science and Technology in Japan are developing a conduction-cooled superconducting combined-function magnet with an elliptical aperture, designed for a compact, rapid-cycling heavy-ion synchrotron~\cite{Japan_Ellipse}.

A two-dimensional planar magnetic field must satisfy Maxwell's equations, and field quality in circular apertures is well understood~\cite{Russenschuck_Design}. 
However, transitioning from conventional circular coil magnets to quasi-polygonal coil magnets requires re-optimization of current distributions to achieve comparable magnetic field properties. 
To date, no comprehensive analytical framework exists for the electromagnetic design of quasi-polygonal bore coil magnets. 
This paper addresses this gap by introducing an analytical method for determining current distributions optimized for quasi-polygonal bores, which is then applied to the design of canted-cosine-theta (CCT) magnets~\cite{CCT_original}. 
The CCT magnet configuration is particularly advantageous due to its ability to produce high-field profiles while maintaining a compact coil structure and low coil stresses~\cite{CCT_thesis}.

The structure of this paper is as follows: \autoref{current_sheet} introduces the relationship between magnetic fields and current distributions in circular apertures, along with the mathematical principles underlying the design. 
We then utilize conformal mapping techniques to extend these results to quasi-polygonal apertures, deriving current distributions that reproduce circular field harmonics in quasi-polygonal geometries. 
In \autoref{App_CCT}, we apply these results to the CCT magnet design, examining the relationship between winding paths and the derived current distributions. 
This analytical approach is then used to design CCT magnets with quasi-polygonal apertures. 
Finally, \autoref{Final} presents a discussion of the results and conclusions.

\section{\label{current_sheet}Current to field relations for a quasi-polygonal current sheet}
\begin{figure*}[!htbp]
	\centering
	\begin{minipage}{0.3\textwidth}
		\includegraphics[height=4.0cm]{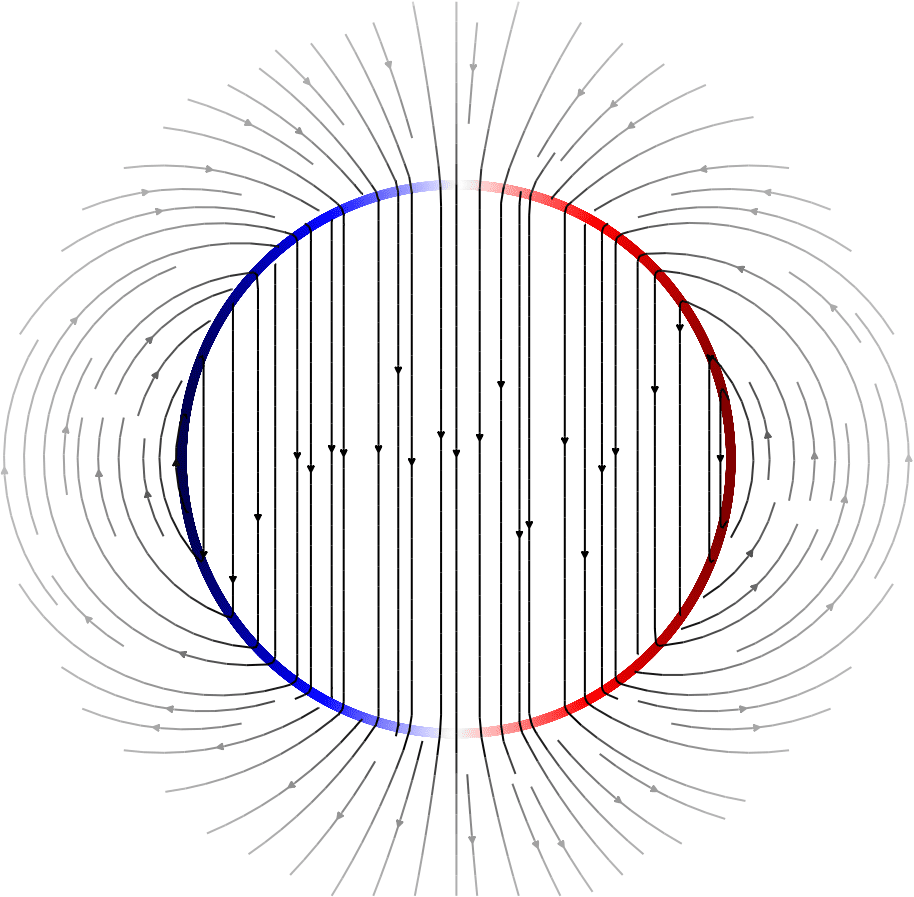}
		\caption*{\(B_1\) dipole}
		\centering
		\label{cir_B1}
	\end{minipage}%
	\hfill
	\begin{minipage}{0.3\textwidth}
		\includegraphics[height=4.0cm]{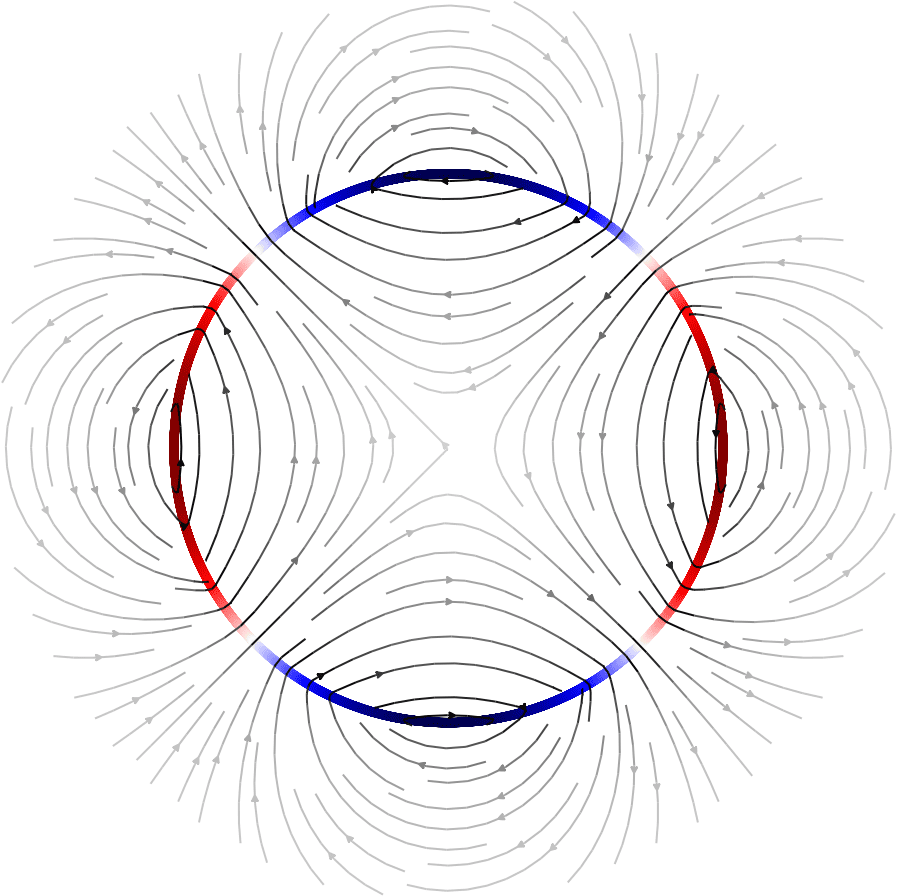}
		\caption*{\(B_2\) quadrupole}
		\centering
		\label{cir_B2}
	\end{minipage}
	\hfill
	\begin{minipage}{0.3\textwidth}
		\includegraphics[height=4.0cm]{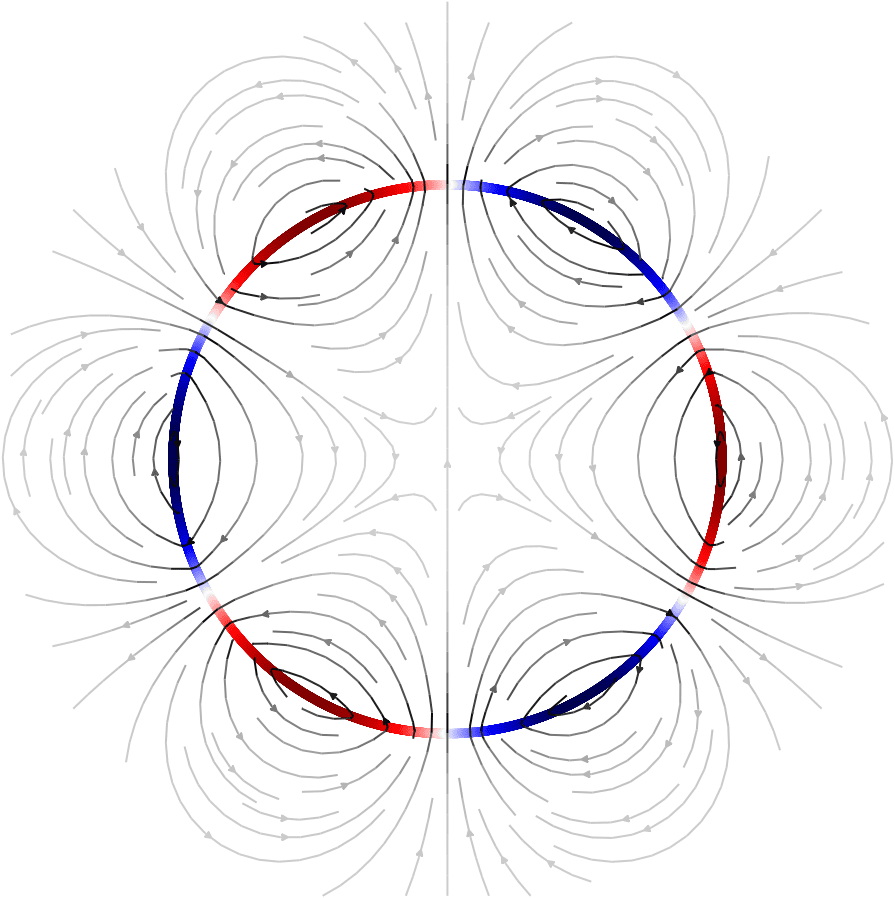}
		\caption*{\(B_3\) sextupole}
		\centering
		\label{cir_B3}
	\end{minipage}
	\caption{The current densities on a circular shell to produce the first 3 circular field harmonics inside an circular aperture.}
	\label{cir_magnetic}
\end{figure*}

For magnets with no axial variation, Maxwell’s equations simplify such that only the axial component of the vector potential, \(A_z\), exists, satisfying Poisson’s equation \(\nabla^2 A_z = -\mu_0 J_z\). This formulation is commonly explored in cylindrical coordinates as follows.
\begin{equation}
	\begin{aligned}
		\vec{B}&=\frac{\partial A_z}{\partial y}\hat{e}_x-\frac{\partial A_z}{\partial x}\hat{e}_y=\frac{1}{\rho}\frac{\partial A_z}{\partial \theta}\hat{e}_{\rho}-\frac{\partial A_z}{\partial \rho}\hat{e}_{\theta},\\
		\nabla^2 A_z &=\frac{\partial^2 A_z}{\partial \rho^2}+\frac{1}{\rho}\frac{\partial A_z}{\partial \rho}+\frac{1}{\rho^2}\frac{\partial^2 A_z}{\partial\theta^2}=-\mu_0J_z.
	\end{aligned}
\end{equation}

The vector potential of a line current carry \(I\) at \(z_0\) in complex plane is expressed as:
\begin{equation}
	A_z\left( z \right) =-\frac{\mu _0I}{2\pi}\Re \left[ \ln \left( \frac{z-z_0}{z_{\text{arbi}}} \right) \right].
\end{equation}
where \(z_{\text{arbi}}\) is an arbitrary constant.
A circular current shell of radius \(\rho_0\) with a \(\cos n\theta\) distribution generates a normal circular harmonic, while a \(\sin n\theta\) distribution produces a skew harmonic, defining the magnetic field's spatial variations and symmetry for different multipole orders~\cite{700_book}.
The vector potentials inside and outside the shell are related through boundary conditions.
The corresponding magnetic potential and current distribution are given as follows:
\begin{equation}
	\begin{aligned}
		\text{normal:\,\,}&A_{z}^{\text{i}}=a_{\text{i}}\rho ^n\cos \left( n\theta \right) ,A_{z}^{\text{o}}=a_{\text{o}}\frac{\cos \left( n\theta \right)}{\rho ^n},\\
		&J_n=-\frac{2a_{\text{o}}}{\mu _0}\frac{n\cos \left( n\theta \right)}{\rho _{0}^{n+1}};\\
		\text{skew:\,\,}&A_{z}^{\text{i}}=a_{\text{i}}\rho ^n\sin \left( n\theta \right) ,A_{z}^{\text{o}}=a_{\text{o}}\frac{\sin \left( n\theta \right)}{\rho ^n},\\
		&J_n=-\frac{2a_{\text{o}}}{\mu _0}\frac{n\sin \left( n\theta \right)}{\rho _{0}^{n+1}};\\
		\text{relation:\,\,}&\frac{a_{\text{i}}}{a_{\text{o}}}=\rho _{0}^{-2n}.
	\end{aligned}
	\label{cir_field}
\end{equation}
Especially,
\begin{equation}
	 J_0=-\frac{a_{\text{o}}}{\mu _0}\frac{1}{\rho _{0}},  A_{z}^{\text{i}}=A_{\text{arbi}}, A_{z}^{\text{o}}=a_{\text{o}}\ln \left( \frac{\rho}{\rho_0} \right)+ A_{\text{arbi}},
	 \label{circular_special}
\end{equation}
\( A_{\text{arbi}} \) is an arbitrary constant.
The \autoref{cir_magnetic} shows the first 3 circular field harmonics by \autoref{cir_field}.

Many physical problems, particularly those involving electrostatics, fluid dynamics, and heat conduction, can be formulated as the determination of a planar potential function subject to specified boundary conditions~\cite{Hassani2013,Riley_Hobson_Bence_2006}. 
Traditionally, these problems are solved using methods such as separation of variables or by applying integral solution formulas.
However, these conventional approaches become impractical when dealing with complex boundary geometries.
Conformal mapping, a mathematical technique, provides a powerful alternative by transforming complex boundary shapes into simpler, more manageable ones, thereby facilitating the solution of the potential problem. 
This method, known as conformal transformation, preserves angles and local geometric properties of the original domain, allowing the problem to be reformulated and solved in a more tractable mathematical space.

In complex analysis, the original 2D plane, denoted as \(z-\)plane, can be described using either Cartesian coordinates \((x, y)\) or polar coordinates \((\rho, \theta)\), where \(J_z\) represents the current density and \(A_z\) represents the magnetic potential.
When a conformal transformation is applied, the transformed \(\zeta-\)plane is described by new Cartesian coordinates \((\eta, \xi)\) and new polar coordinates \((P, \Theta)\), with corresponding current density \(\mathbb{J}_z\) and magnetic potential  \(\mathbb{A}_z\).

If the new independent variable \(\zeta\) is an analytic function of the original variable \(z\) within the domain of interest as \(\zeta=\zeta\left(z\right)\), and the transformation adheres to the Cauchy-Riemann conditions, then the 2D Poisson equation in the original \(z-\)plane transforms into an equivalent 2D Poisson equation in the transformed \(\zeta-\)plane.
However, the 2-D source term's intensity undergoes rescaling as follows:
\begin{equation}
	\begin{aligned}
		& u_{xx} + u_{yy} = f(x, y), \Longrightarrow \\
		& u_{\eta \eta} + u_{\xi \xi} = \frac{1}{|\zeta^\prime(z)|^2} f\left[x(\xi, \eta), y(\xi, \eta)\right] .
	\end{aligned}
\end{equation}
Besides, if the source is a Dirac Delta function, the scaling factor is 1. 
If the source is a one-dimensional line distribution, the scaling factor is given by \(\frac{1}{|\zeta^\prime(z)|}\)
, which is the specific case considered in this paper.

\subsection{Elliptical shell}
Here ,we consider a conformal mapping widely applied in engineering~\cite{ivanov1994handbook}. The Zhukovskii transformation of any complex number \(z\) to \(\zeta\) is given by:
\begin{equation}
	\zeta \left( z \right) =\frac{1}{2}\left( z+\frac{c^2}{z} \right).
\end{equation}
Here, \(c\) represents the focal point of the transformed ellipses.
This transformation is fundamental in modeling airflow around airfoils and analyzing potential flow in fluid dynamics.
In detail, the Zhukovskii mapping maps a circle in the \(z\)-plane to an ellipse or line in the \(\zeta\)-plane.
The new Cartesian coordinates \((\eta, \xi)\) in the \(\zeta\)-plane can be written as follows:
\begin{equation}
	\begin{aligned}
		\left\{ \begin{array}{l}
			\eta =\frac{1}{2}\left( x+\frac{c^2x}{x^2+y^2} \right) =\frac{1}{2}\left( \rho +\frac{c^2}{\rho} \right) \cos \theta ,\\
			\xi =\frac{1}{2}\left( y-\frac{c^2y}{x^2+y^2} \right) =\frac{1}{2}\left( \rho -\frac{c^2}{\rho} \right) \sin \theta .\\
		\end{array} \right.
	\end{aligned}
\end{equation}
From this, we can derive the semi-axes of the mapped ellipse in the \(\zeta\)-plane, which originates from a circle \(\rho\) in the \(z\)-plane, as follows.
\begin{equation}
\left\{
\begin{aligned}
	& a = \frac{1}{2} \left( \rho + \frac{c^2}{\rho} \right), \quad 
	b = \frac{1}{2} \left( \rho - \frac{c^2}{\rho} \right), \\[8pt]
	& \eta = a \cos \theta, \quad 
	\xi = b \cos \theta, \\[8pt]
	& \frac{\eta^2}{a^2} + \frac{\xi^2}{b^2} = 1, \quad 
	a^2 - b^2 = c^2.
\end{aligned}
\right.
\end{equation}
The set of mapped ellipses all share a common focal point \(c\). 
As the radius \(\rho\) of the original circle increases from the boundary circle \(\rho = c\) to infinity in the \(z\)-plane, both the major axis \(a\) and minor axis \(b\) of the corresponding ellipses expand without bound.
This transformation maps the exterior of the circle \(\rho = c\) in the \(z\)-plane onto the entire \(\zeta\)-plane, stretching it from a line segment between \(-c\) and \(+c\) on the real axis, into large ellipses with foci at \(\pm c\).
Similarly, as \(\rho\) decreases from the boundary circle \(\rho = c\) to zero, \(a\) and \(b\) also expand infinitely, causing the interior of the circle \(\rho = c\) in the \(z\)-plane to map onto the entire \(\zeta\)-plane.

\begin{figure}[htbp]
	\centering
	\includegraphics[width=7.5cm]{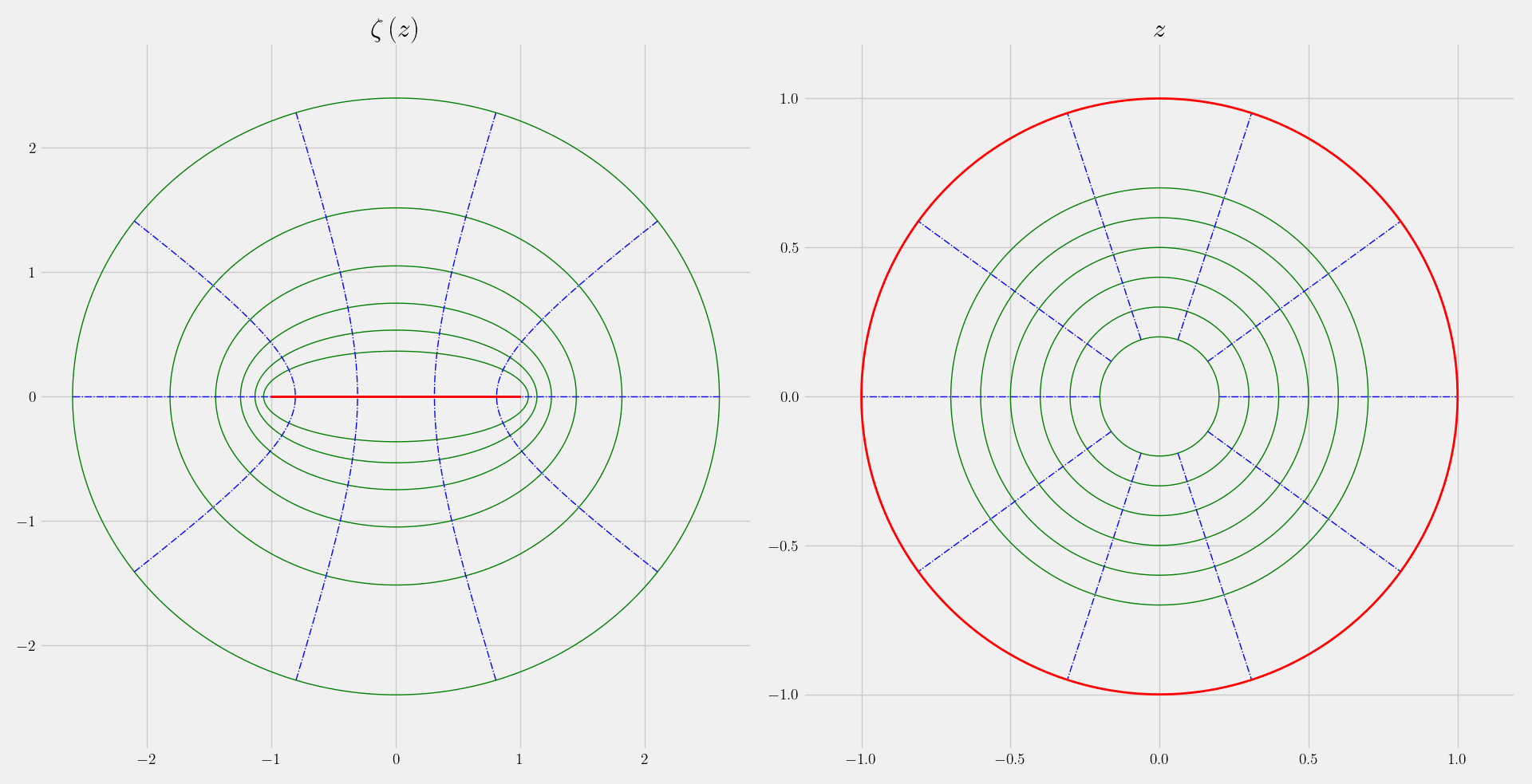}
	\caption{The Zhukovskii transformation maps a set of circles in the \(z\)-plane into a set of ellipses in the \(\zeta\)-plane. As the circle shrinks from \(\rho = c\) in the right image, the corresponding ellipse expands outward from the central axis into infinity in the left image.}
	\label{map_ellipse}
\end{figure}

The inverse mapping of Zhukovskii mapping is given by:
\begin{equation}
	z = \zeta \pm \sqrt{\zeta^2 - c^2},
\end{equation}
where the \(+\) sign corresponds to the exterior of the boundary circle \(\rho = c\) in the \(z\)-plane, and the \(-\) sign corresponds to the interior of the boundary circle \(\rho = c\) in the \(z\)-plane. 
In the following analysis, we focus on the negative sign in the inverse transformation, which maps the constrained domain of definition around the origin in the \(z\)-plane onto the entire domain of the \(\zeta\)-plane.
The corresponding geometric representation is illustrated in the \autoref{map_ellipse}. 

\begin{figure}[htbp]
	\centering
	\includegraphics[width=7.5cm]{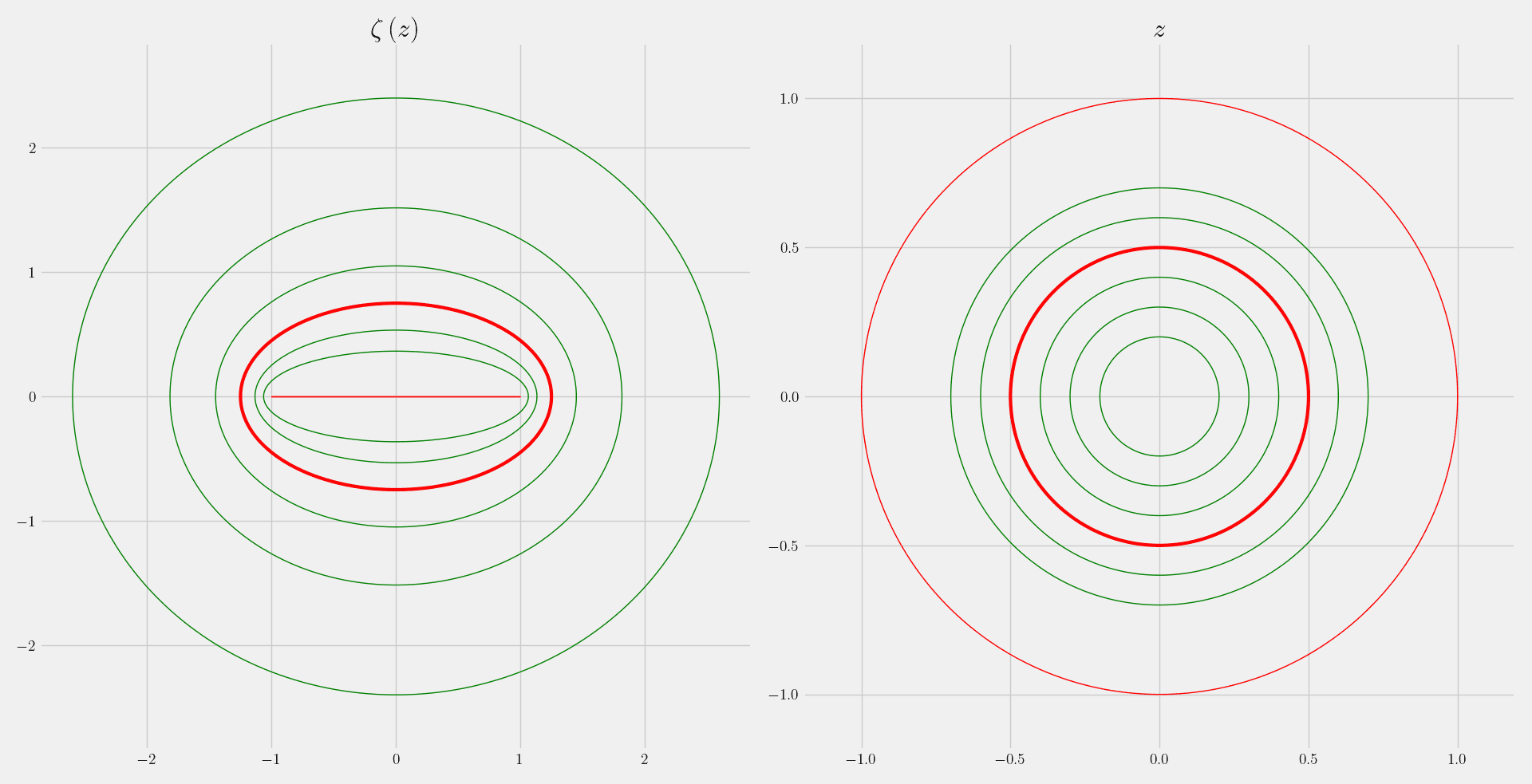}
	\caption{The thick red circle in the right image (thick red ellipse in the left image) represents the current shell, while the outer light red circle in the right image indicates the domain boundary in the \(z\)-plane.}
	\label{map_ellipse_shell}
\end{figure}

Considering an elliptical shell carrying current in the \(\zeta\)-plane, it corresponds to a circular shell with radius \(\rho_0<c\) in the \(z\)-plane through Zhukovskii mapping, which carries the corresponding current. 
Our objective is to derive the current distribution that generates the magnetic potential inside the elliptical shell, given by \(\mathbb{A}_{z}^{\text{i}} = a_{\text{i}}P^n\cos(n\Theta)\), in the \(\zeta\)-plane.
By using the inverse Zhukovskii transformation, the corresponding magnetic potential outside the current shell \(\rho_0\) but inside boundary circle \(c\) in the \(z\)-plane is given by \(A_{z}^{\text{o}} = \mathbb{A}_{z}^{\text{i}}\), as shown in \autoref{map_ellipse_shell}.
And the intrinsic boundary condition for this exterior magnetic potential \(A_{z}^{\text{o}}\) in the \(z\)-plane is automatically satisfied, as \(A_{z}^{\text{o}}\) totally comes from \(\mathbb{A}_{z}^{\text{i}}\).
The transformation of the potential between the coordinates of the \(\zeta\)- and \(z\)-planes is expressed as:
\begin{equation}
	\begin{aligned}
	A_{z}^{\text{o}} = \mathbb{A}_{z}^{\text{i}} & = a_i P^n\cos(n\Theta) \\
	 & = a_i \Re \left\{ \frac{1}{2^n} \left( \rho e^{i\theta} + \frac{c^2}{\rho}e^{-i\theta} \right)^n \right\}.
	\end{aligned}
\end{equation}

\begin{figure*}[!htbp]
	\centering
	\begin{minipage}{0.3\textwidth}
		\includegraphics[height=4.0cm]{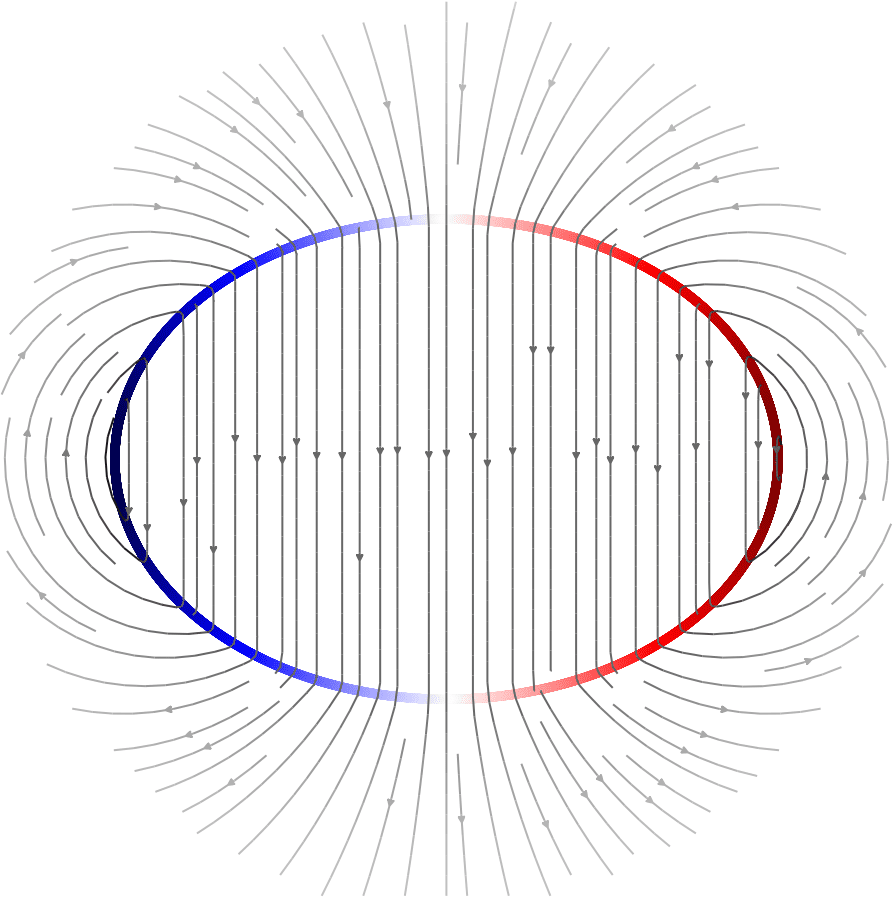}
		\centering
		\caption*{\(B_1\) dipole}
		\label{ellipse_B1}
	\end{minipage}
	\hfill
	\begin{minipage}{0.3\textwidth}
		\includegraphics[height=4.0cm]{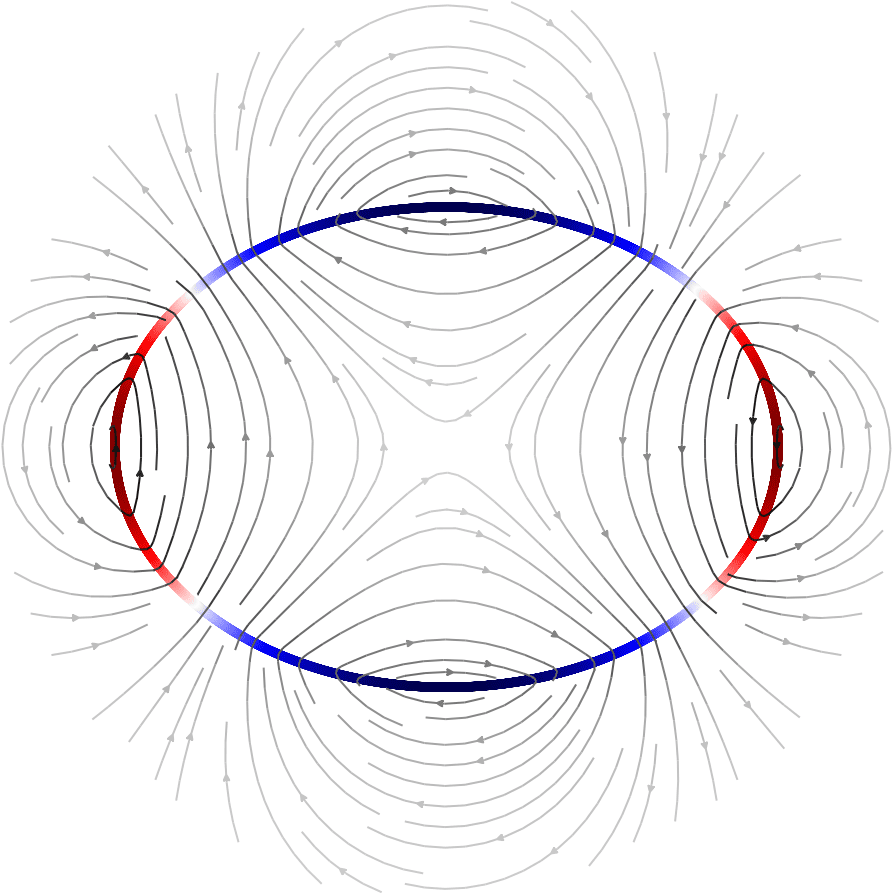}
		\centering
		\caption*{\(B_2\) quadrupole}
		\label{ellipse_B2}
	\end{minipage}
	\hfill
	\begin{minipage}{0.3\textwidth}
		\includegraphics[height=4.0cm]{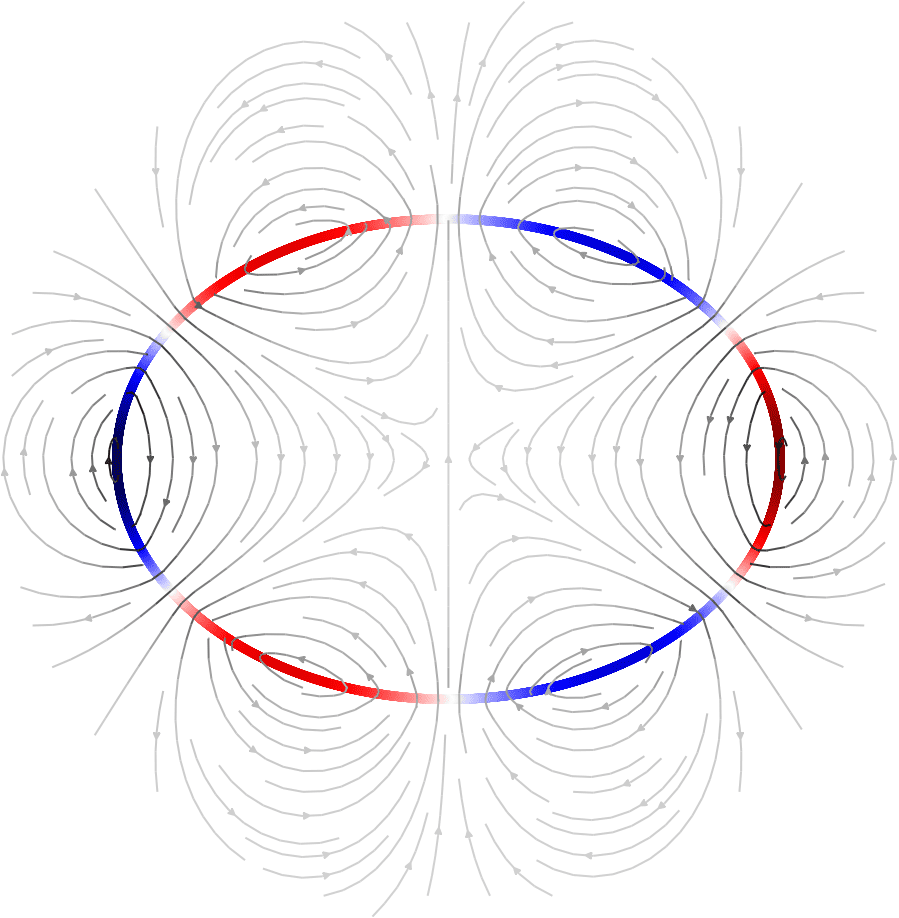}
		\centering
		\caption*{\(B_3\) sextupole}
		\label{ellipse_B3}
	\end{minipage}
	\caption{The current densities on an elliptical shell to produce the first 3 circular field harmonics inside an elliptical aperture.}
	\label{ellipse_magnetic}
\end{figure*}

Given the exterior magnetic potential \(A_{z}^{\text{o}}\) in the \(z\)-plane and applying the previously discussed current-potential relationship, the current density on the circular shell at \(\rho_0\) in the \(z\)-plane can be determined.
The magnetic potential in the constrained domain of the \(z\)-plane consists of two components: one originating from the current on the shell at \(\rho_0\), represented as a piecewise function as mentioned in \autoref{cir_field}, and the other as a continuous function involving polynomials of \(\rho\).
By mapping these current distribution back to the \(\zeta\)-plane, the corresponding current density distribution \(\mathbb{J}_{z}\) for the elliptical configuration is obtained.
The current distribution and the generation of circular field harmonics inside the elliptical shell are illustrated in \autoref{ellipse_magnetic}.

A particularly interesting case occurs when the current shell in the \(\zeta\)-plane carries a non-zero net current \(\mathbb{J}_{z} \sim \frac{1}{|\zeta^\prime(z)|} \), leading to a zero magnetic field inside the shell and a magnetic field outside that decays slowly as \(1/P\).
This situation can also be described in the \(z\)-plane, where the magnetic potential consists of two components: one due to the current on the shell \(J_{z} = \text{const}\) at \(\rho_0\), as illustrated in \autoref{circular_special}, and an additional term involving \(-\ln \rho\) that cancels the previous one. 
This additional term ensures that, in the \(z\)-plane, \(A_{z}^{\text{i}}(\rho = 0) = \infty\) and \(A_{z}^{\text{o}}(\rho = \infty) = 0\).
This is consistent with the corresponding behavior in the \(\zeta\)-plane, where \(\mathbb{A}_{z}^{\text{o}}(P = \infty) = \infty\) and \(\mathbb{A}_{z}^{\text{i}}(P = 0) = 0\).

For instance, to generate a normal sextupole magnetic field \(\mathbb{A}_{z}^{\text{i}} = a_{\text{i}}P^3\cos(3\Theta)\) inside an elliptical current shell, the corresponding magnetic potential outside the shell at \(\rho_0\) in the \(z\)-plane is given by:
\begin{equation}
	\begin{aligned}
		& A_{z}^{\text{o}} = a_{\text{i}}P^3\cos(3\Theta)= \Re \left\{ \frac{a_{\text{i}}}{2^3} \left( \rho e^{i\theta} + \frac{c^2}{\rho}e^{-i\theta} \right)^3 \right\}=\\
		&\frac{a_{\text{i}}}{2^3}\left[ \left( \rho ^3\cos 3\theta +3\rho c^2\cos \theta \right) +\left( \frac{c^6}{\rho ^3}\cos 3\theta +3\frac{c^4}{\rho}\cos \theta \right) \right] .
	\end{aligned}
\end{equation}
The terms with negative exponents of \(\rho\) indicate the current density on the circular shell at \(\rho_0\) in the \(z\)-plane by \autoref{cir_field}, expressed as:
\begin{equation}
	J(\theta)=-\frac{2}{\mu_0}\frac{a_{\text{i}}}{2^3} \left( \frac{3c^6}{\rho_0 ^4}\cos 3\theta +\frac{3c^4}{\rho_0^2}\cos \theta \right).
\end{equation}
The potential inside the current shell in the \(z\)-plane can be expressed as a sum of polynomials in \(\rho\):
\begin{equation}
	\begin{aligned}
		& \mathbb{A}_{z}^{\text{o}} = A_{z}^{\text{i}} = \frac{a_{\text{i}}}{2^3} \times \\
		& \left[ \left( \rho^3 \cos 3\theta + 3\rho c^2 \cos \theta \right) +  \left( \frac{c^6}{\rho_0^6} \rho^3 \cos 3\theta + 3\frac{c^4}{\rho_0^2} \rho \cos \theta \right) \right].
	\end{aligned}
	\label{uninterest}
\end{equation}
We observe that:
\begin{equation}
	\begin{aligned}
		a &= \frac{1}{2}\left( \rho_0 + \frac{c^2}{\rho_0} \right), \quad |b| = \frac{1}{2}\left( \frac{c^2}{\rho_0} - \rho_0 \right); \\
		|\zeta'(z)| &= \left| \frac{1}{2}\left( 1 - \frac{c^2}{z^2} \right) \right| 
		= \frac{1}{\rho_0}\sqrt{\frac{1}{4}\left( \rho_0^2 + \frac{c^4}{\rho_0^2} - 2c^2 \cos 2\theta \right)} \\
		&= \frac{1}{\rho_0} \sqrt{a^2 \sin^2 \theta + b^2 \cos^2 \theta}; \\
		&\frac{3c^4}{\rho_0^2} \bigg/ \frac{3c^6}{\rho_0^4} = \frac{\rho_0^2}{c^2} = \frac{a - |b|}{a + |b|},
	\end{aligned}
\end{equation}

The corresponding current density of the elliptical shell in the $\zeta$-plane is given by:
\begin{equation}
	\begin{aligned}
		\mathbb{J}_{z}&=\frac{J(\theta)}{|\zeta^\prime \left( z \right)|}=\frac{J(\theta)\rho_0}{\sqrt{a^2\sin ^2\theta +b^2\cos ^2\theta}}\\
		&\sim \frac{\cos 3 \theta + \frac{a-|b|}{a+|b|} \cos \theta}{\sqrt{a^2\sin ^2\theta +b^2\cos ^2\theta}}.
	\end{aligned}
\end{equation}

This result aligns with findings in the literature~\cite{schnizer2017advanced}, which provides intrinsic solutions in elliptic cylindrical coordinates. 
Although the exterior domain in the \(\zeta\)-plane for \autoref{uninterest} is not the primary focus of our analysis, its magnetic potential can also be determined using a similar method.
The magnetic potential in the interior domain of the \(z\)-plane, \(A_{z}^{\text{i}} = \mathbb{A}_{z}^{\text{o}} \), can be expressed in terms of positive powers of \(\rho\), consistent with the observation that the magnetic potential at the origin \(\rho=0\) in the \(z\)-plane is zero, corresponding to the zero magnetic potential at infinity in the \(\zeta\)-plane when the net current on the shell is zero. 
However, expressing the coordinates \(z = (\rho, \theta)\) in terms of \(\zeta = (P, \Theta)\) in \autoref{uninterest} is not straightforward, complicating direct calculations.

\subsection{Quasi-polygonal shell}

\begin{figure}[htbp]
	\centering
	\includegraphics[width=8.1cm]{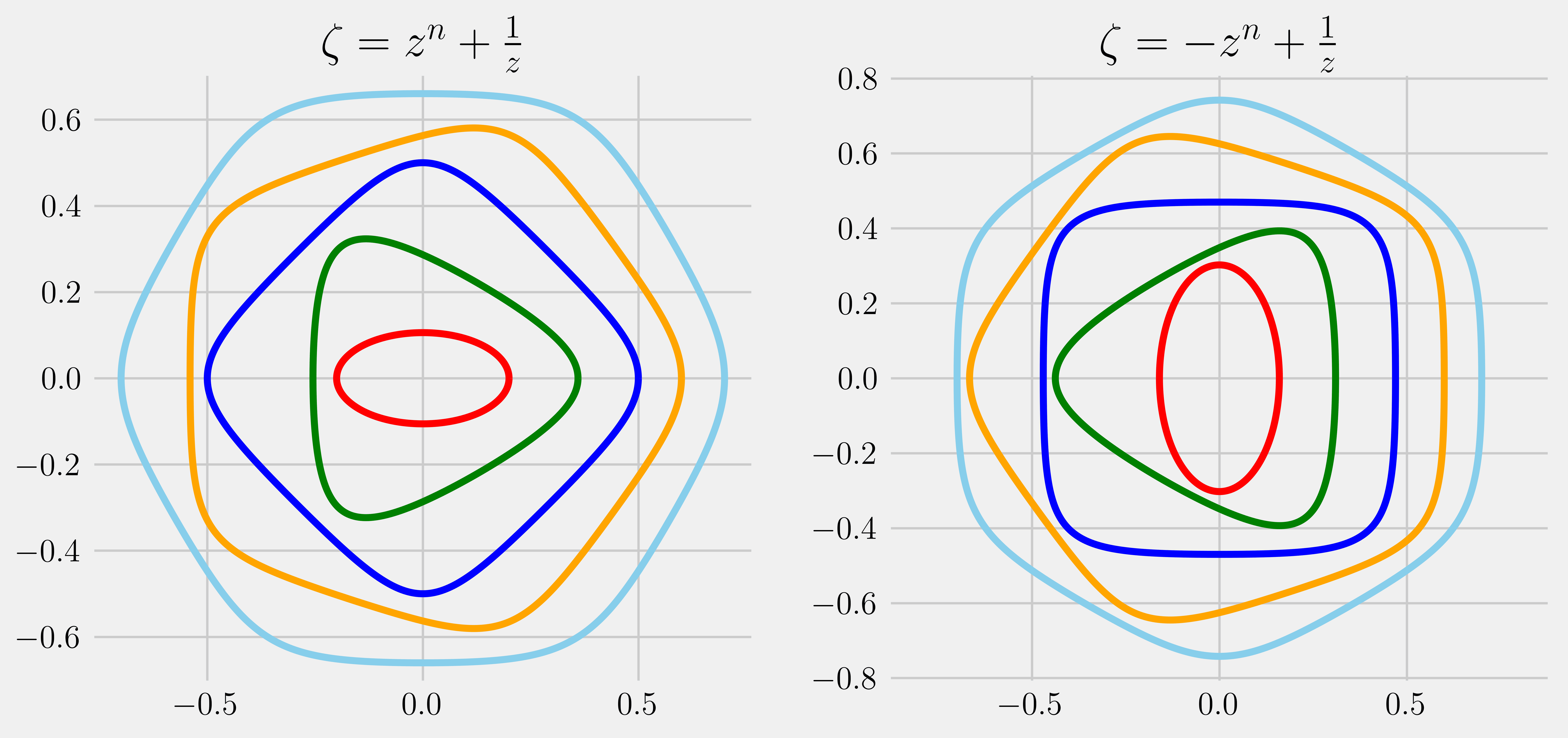}
	\caption{A set of transformations resulting in quasi-polygonal shells: the left one is defined by \(\zeta = \frac{z^n}{c^{n-1}}+ \frac{c^2}{z}\), and the right one by \(\zeta = -\frac{z^n}{c^{n-1}}+ \frac{c^2}{z} \). The curves are scaled to display all shapes on one graph.}
	\label{transform}
\end{figure}

Similar to the Zhukovskii mapping, other conformal mappings can transform circles in the \(z\)-plane into various geometries in the \(\zeta\)-plane, providing an analytical method to derive circular field harmonics within the corresponding apertures in the \(\zeta\)-plane.
We present a set of transformations that generate quasi-polygonal shells, mapping the constrained domain of definition around the origin in the \(z\)-plane to the entire domain in the \(\zeta\)-plane, as shown in \autoref{transform}. 
We propose two mappings: \(\zeta = \frac{z^n}{c^{n-1}}+ \frac{c^2}{z}\), which forms a quasi-\(n\)-polygonal shell, and \(\zeta = -\frac{z^n}{c^{n-1}} + \frac{c^2}{z}\), resulting in a rotated quasi-\(n\)-polygonal shell.
We abandon the factor \(\frac{1}{2}\) of the Zhukovskii mapping.
The parameter \(c\) governs the shape of the mapping and serves as a dimensional adjustment factor, typically set to 1 by default.
The inverse mapping from \(\zeta\) to \(z\) is complex and does not yield a unique solution.
Therefore, selecting an appropriate constrained domain of definition in the \(z\)-plane is essential to ensure the inverse mapping provides a unique solution.
Notably, the Zhukovskii mapping can be considered a special case of quasi-polygonal mapping, specifically with two sides. 
In this paper, we present two examples: a quasi-triangular shell and a quasi-square shell.

\begin{figure}[!htbp]
	\centering
	\includegraphics[width=7.5cm]{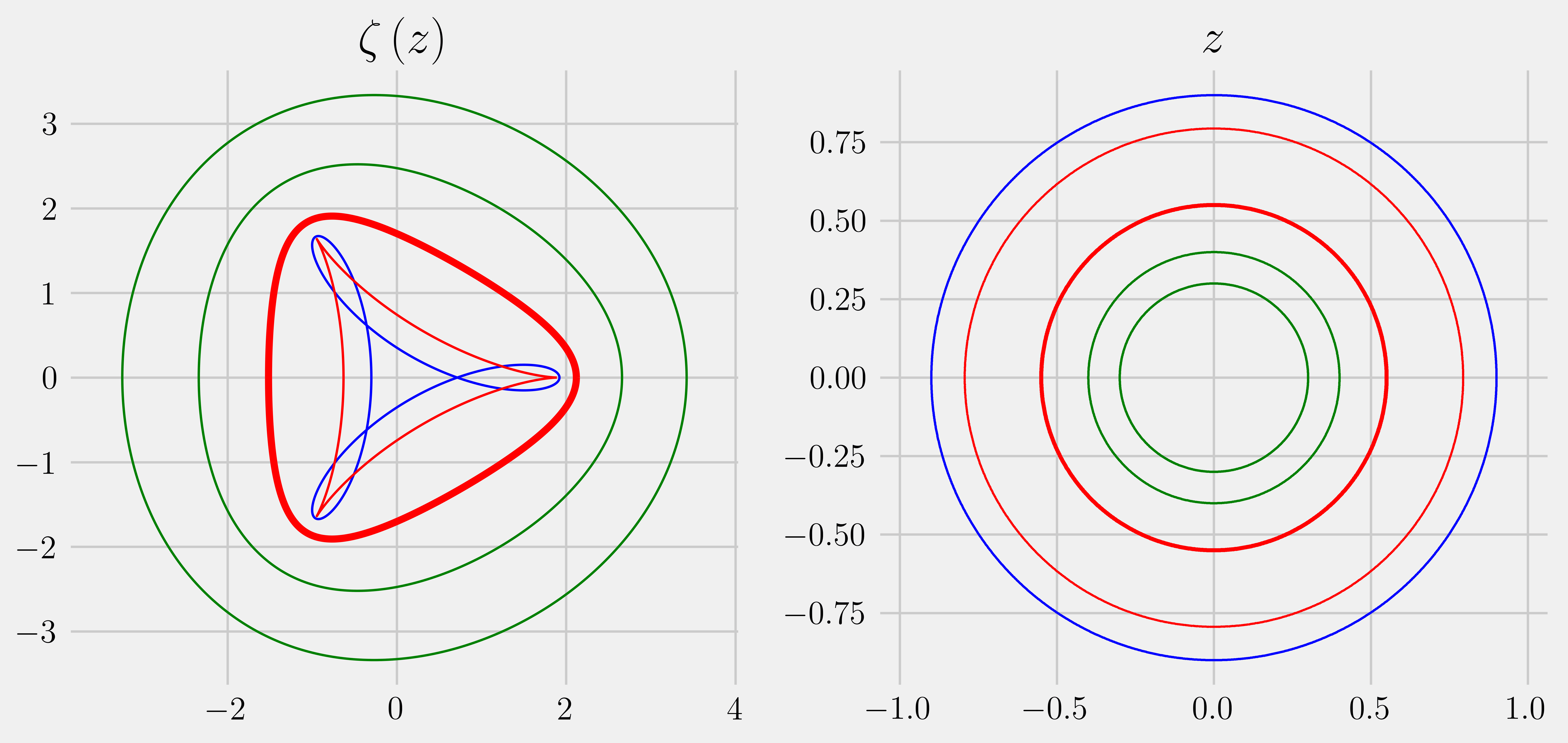}
	\caption{\label{map_triangle_knotted}The outer blue circle in the right image corresponds to the inner blue closed knotted curves in the left image, while the inner green circle in the right image transforms into the outer green quasi-triangle in the left image.}
\end{figure}

\begin{figure}[!htbp]
	\centering
	\includegraphics[width=7.5cm]{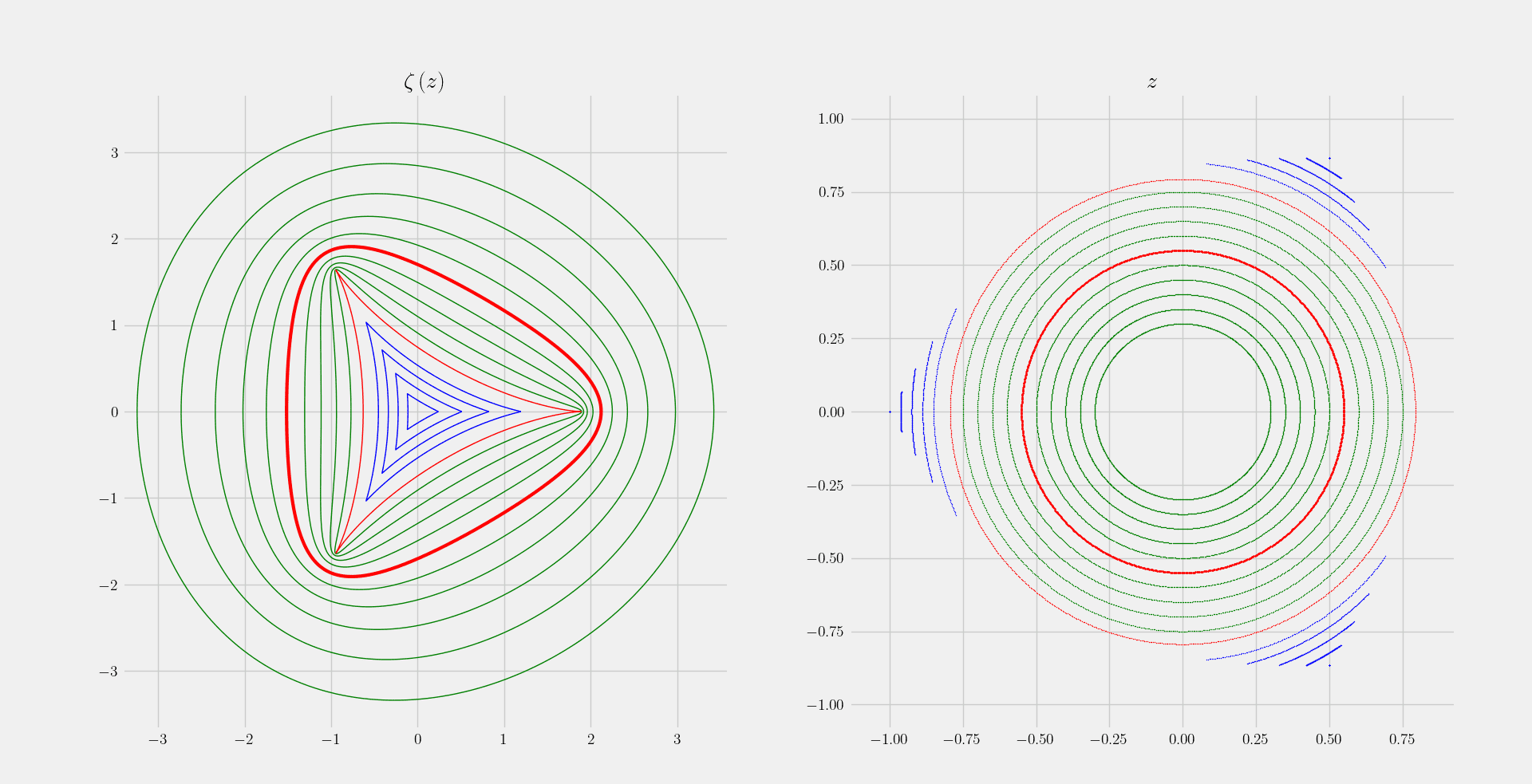}
	\caption{The thick red circle in the right image (thick red quasi-triangle in the left image) represents the current shell, while the outer light red circle \(\rho = \frac{c}{2^{1/3}}\) marks the boundary between closed knotless curves and closed knotted curves. The constrained domain of definition in the right image corresponds to the entire domain in the left image through the transformation.}
	\label{map_triangle_shell}
\end{figure}

\begin{figure*}[!htbp]
	\centering
	\begin{minipage}{0.30\textwidth}
		\includegraphics[height=4.0cm]{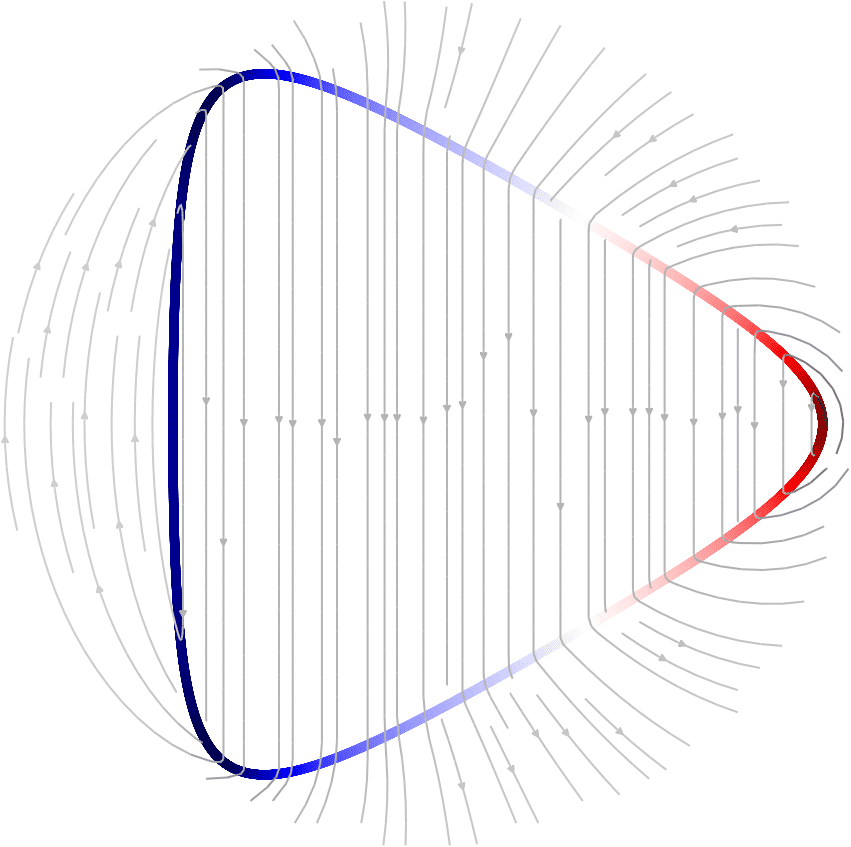}
		\caption*{\(B_1\) dipole}
		\label{triangular_B1}
	\end{minipage}%
	\hfill
	\begin{minipage}{0.30\textwidth}
		\includegraphics[height=4.0cm]{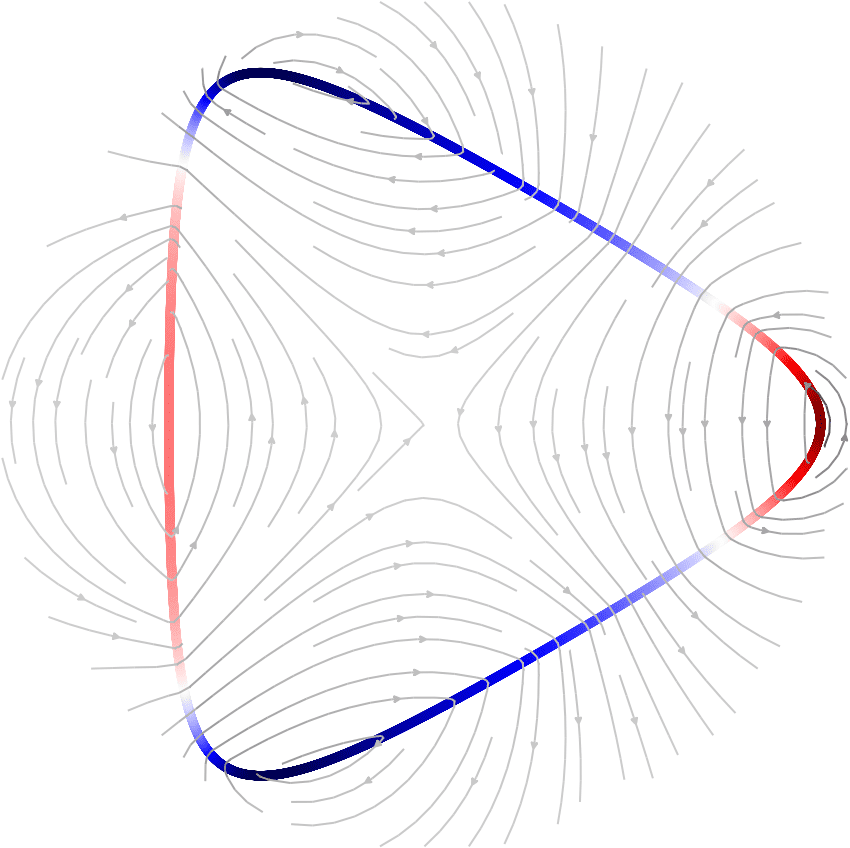}
		\caption*{\(B_2\) quadrupole}
		\label{triangular_B2}
	\end{minipage}
	\hfill
	\begin{minipage}{0.30\textwidth}
		\includegraphics[height=4.0cm]{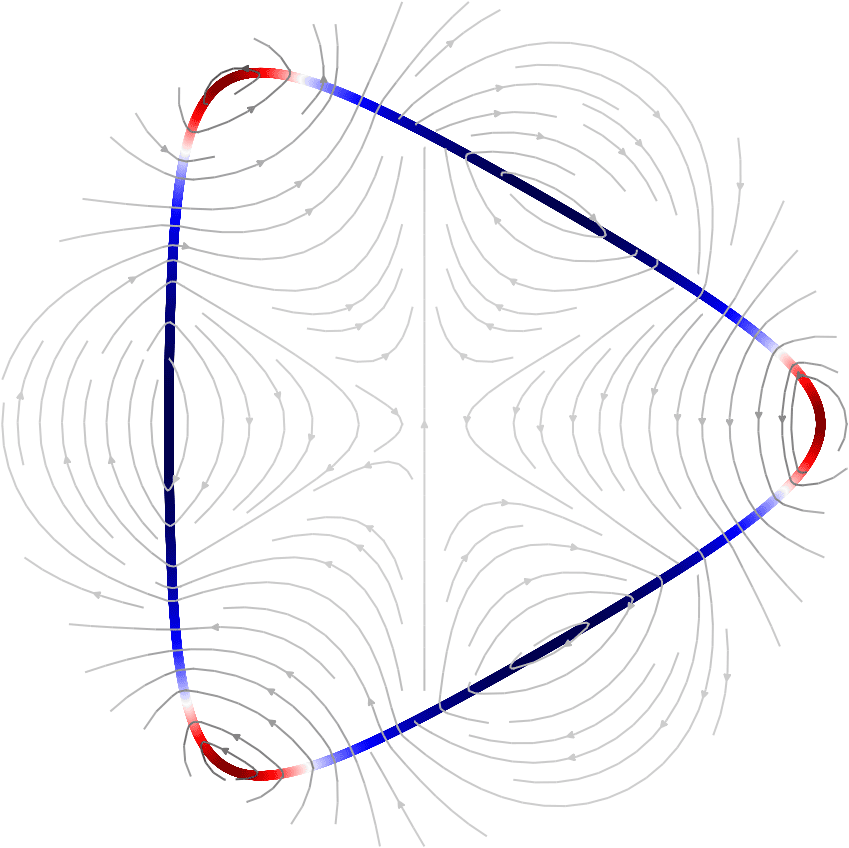}
		\caption*{\(B_3\) sextupole}
		\label{triangular_B3}
	\end{minipage}
	\caption{The current densities on a quasi-triangular shell to produce the first 3 circular field harmonics inside a quasi-triangular aperture.}
	\label{triangular_magnetic}
\end{figure*}

The transformation to a triangle form is given by:
\begin{equation}
	\zeta \left( z \right) =\frac{z^2}{c}+\frac{c^2}{z}.
\end{equation}
As the circle in the \(z\)-plane expands outward from the origin \(\rho=0\), the corresponding closed curve in the \(\zeta\)-plane contracts inward from infinity.
The shape of this closed curve in the \(\zeta\)-plane resembles a quasi-triangle. 
When \(\rho\) increases beyond \(\frac{c}{2^{1/3}}\), the corresponding curve in the \(\zeta\)-plane becomes self-intersecting and have some knotted segments, as illustrated in \autoref{map_triangle_knotted}. 
To prevent this self-intersection, it is necessary to remove certain arcs from the circle \(\rho\), keeping the mapped curve in \(\zeta\)-plane closed and non-self-intersecting.
Ultimately, when \(\rho\) reaches \(c\), the corresponding curve reduces to a single point \(P=0\) in the \(\zeta\)-plane.
And the corresponding mapping from constrained domain of definition in the \(z\)-plane into the whole domain in the \(\zeta\)-plane can be in the  \autoref{map_triangle_shell}.
Here are the formulas associated with the domain that removes certain arcs.
\begin{equation}
	\begin{aligned}
		\frac{c}{2^{1/3}} & <\rho <c, \Delta \theta =\arccos \frac{c^3}{2\rho ^3},\\
		\theta \in &\left[ \Delta \theta ,\frac{2\pi}{3}-\Delta \theta \right] \cup \left[ \frac{2\pi}{3}+\Delta \theta ,\frac{4\pi}{3}-\Delta \theta \right] \\
		&\cup \left[ \frac{4\pi}{3}+\Delta \theta ,2\pi -\Delta \theta \right].
	\end{aligned}
\end{equation}

The following expressions describe the potential in both the inner \(\zeta\)-plane \(\mathbb{A}_z^\text{i}\)  and outer \(z\)-plane \(A_z^\text{o}\) for dipole, quadrupole, and sextupole components:
\begin{equation}
	\begin{aligned}
	\mathbb{A}_z^\text{i} =& A_z^\text{o}, \\
	P\cos \Theta =& -\frac{\rho^3}{c^2}\cos 3\theta + \frac{c^2}{\rho}\cos \theta,\\
	P^2\cos 2\Theta =& \frac{\rho^6}{c^4}\cos 6\theta - 2\rho^2\cos 2\theta + \frac{c^4}{\rho^2}\cos 2\theta,\\
	P^3\cos 3\Theta =& -\frac{\rho^9}{c^6}\cos 9\theta + 3\frac{\rho^5}{c^2}\cos 5\theta\\
	& - 3\rho c^2\cos \theta + \frac{c^6}{\rho^3}\cos 3\theta.
	\end{aligned}
\end{equation}
Based on these expressions, the current distribution in a quasi-triangular shell for the corresponding circular field harmonics inside is given by:
\begin{equation}
	\mathbb{J}_z \sim \frac{\cos n\theta}{|\zeta'(z)|}, \quad \text{for} \, n = 1, 2, 3.
\end{equation}
There are no additional cosine terms for these corresponding circular field harmonics, but higher-order circular field harmonics necessitate the inclusion of additional cosine terms.
For a quasi-triangular shell, the current distribution, the resulting circular field harmonics inside, and the corresponding field outside are illustrated in \autoref{triangular_magnetic}.
We choose a smaller \(\rho_0\) to ensure the mapped quasi-triangle is convex.

\begin{figure}[!htbp]
	\centering
	\includegraphics[width=7.5cm]{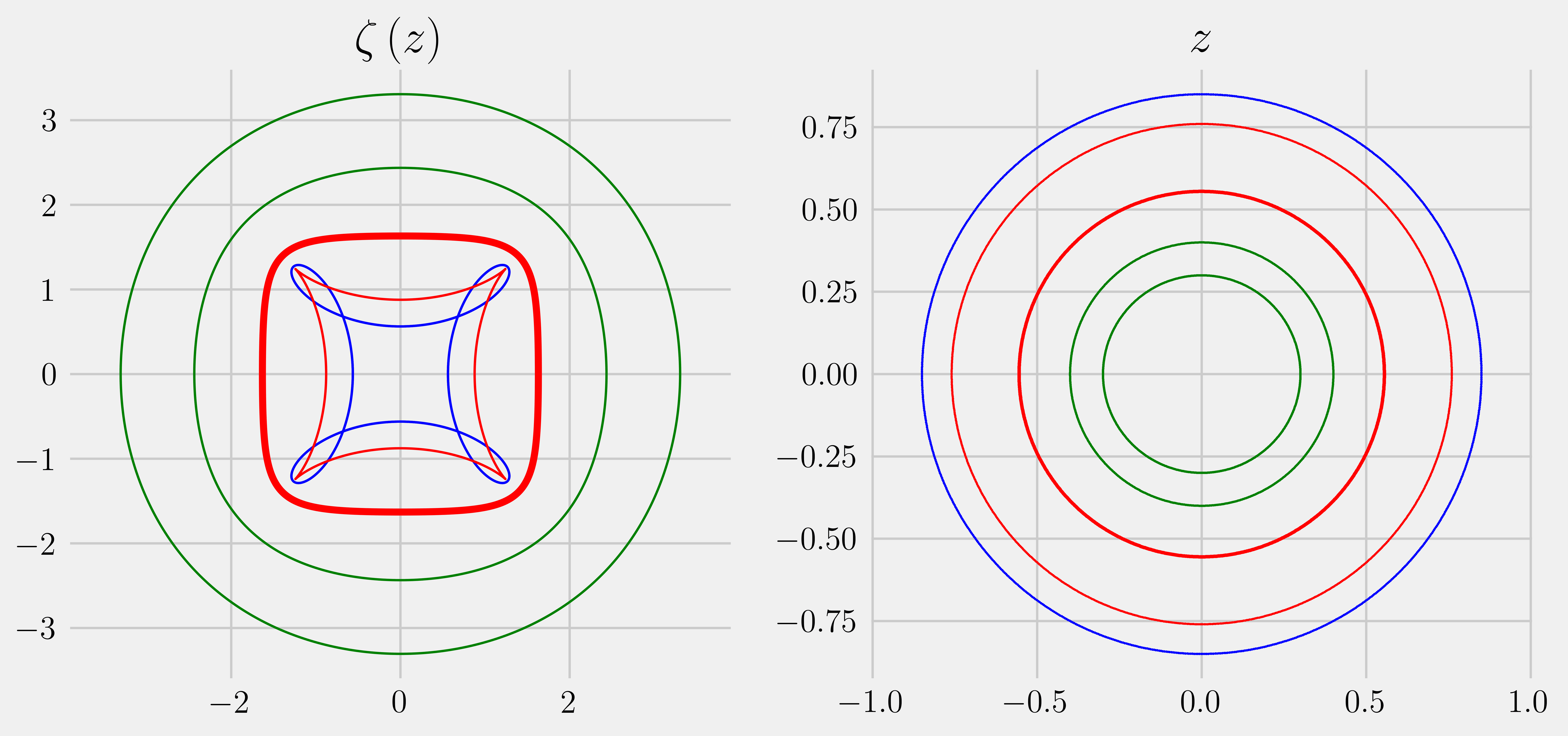}
	\caption{The outer blue circle in the right image corresponds to the inner blue closed knotted curves in the left image, while the inner green circle in the right image transforms into the outer green quasi-square in the left image.}
	\label{map_square_knotted}
\end{figure}
\begin{figure}[!htbp]
	\centering
	\includegraphics[width=7.5cm]{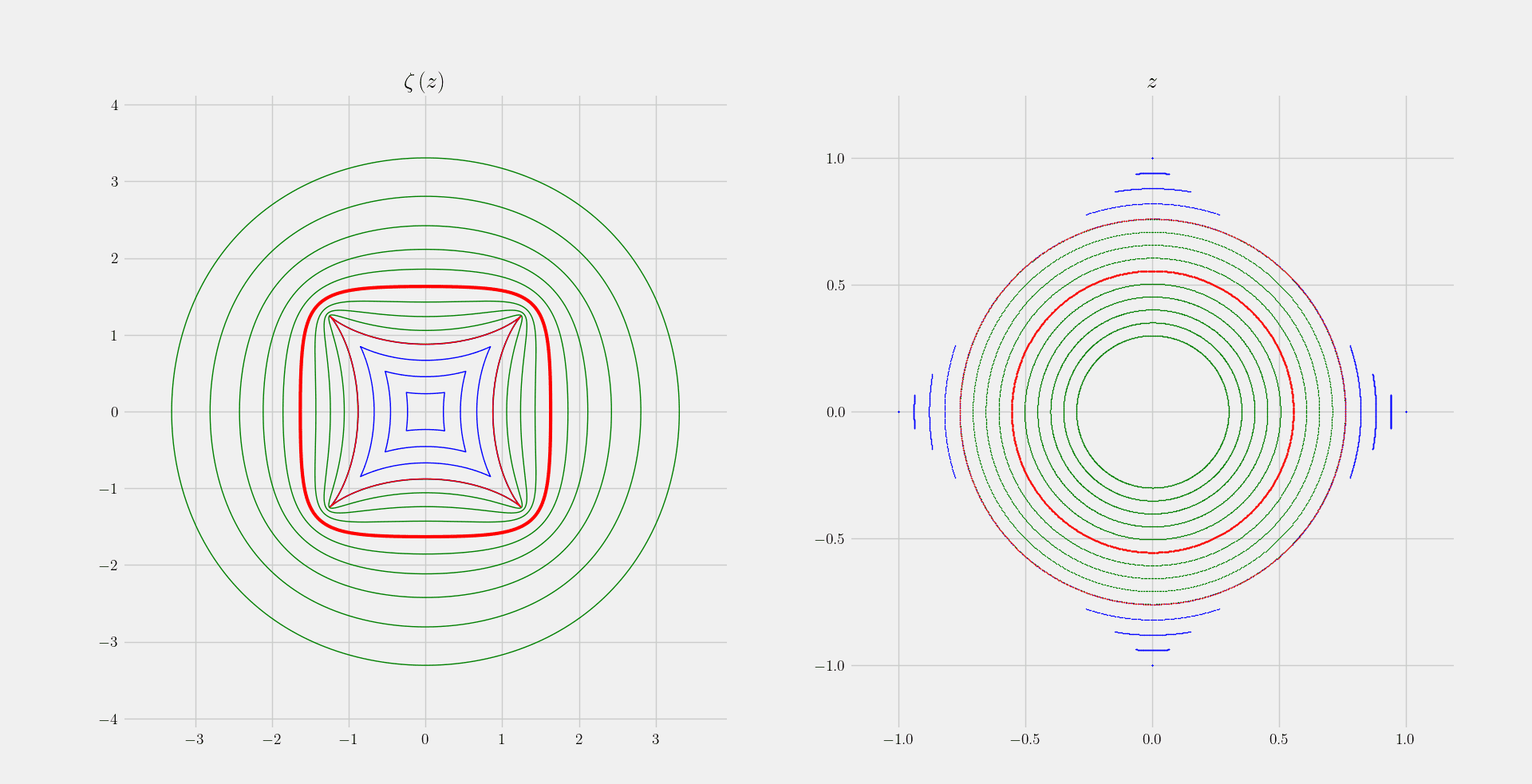}
	\caption{The thick red circle in the right image (thick red quasi-square in the left image) represents the current shell, while the outer light red circle \( \rho = \frac{c}{3^{1/4}} \) marks the boundary between closed knotless curves and closed knotted curves. The constrained domain of definition in the right image corresponds to the entire domain in the left image through the transformation.}
	\label{map_square_shell}
\end{figure}
\begin{figure*}[!htbp]
	\centering
	\begin{minipage}{0.3\textwidth}
		\includegraphics[height=4.0cm]{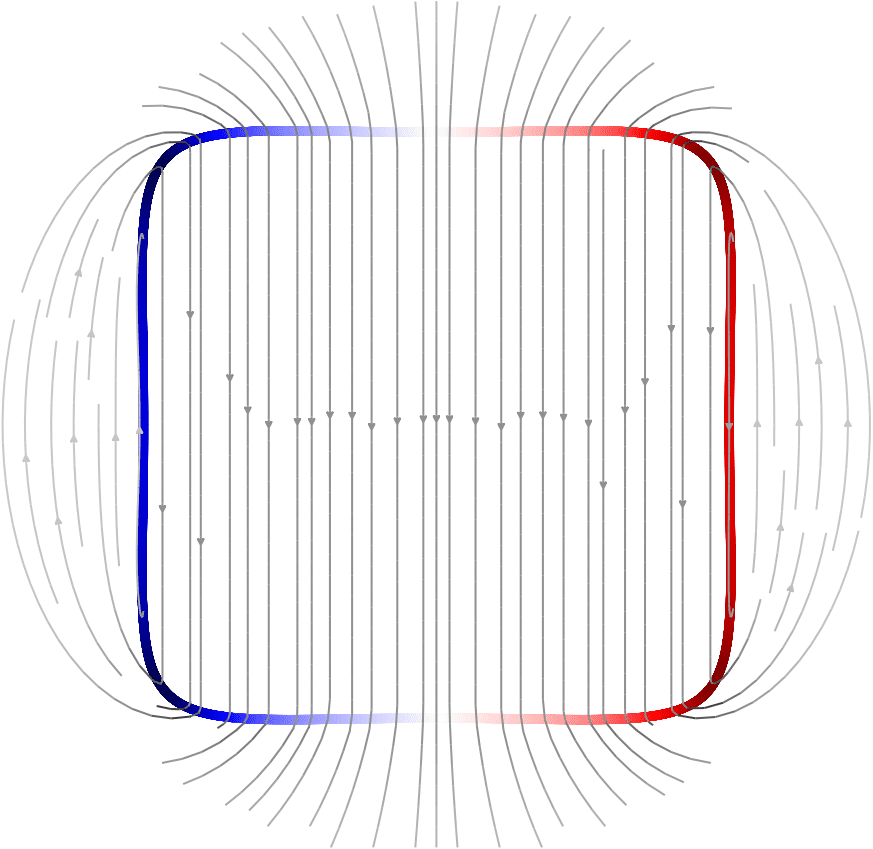}
		\caption*{\(B_1\) dipole}
		\label{square_B1}
	\end{minipage}%
	\hfill
	\begin{minipage}{0.3\textwidth}
		\includegraphics[height=4.0cm]{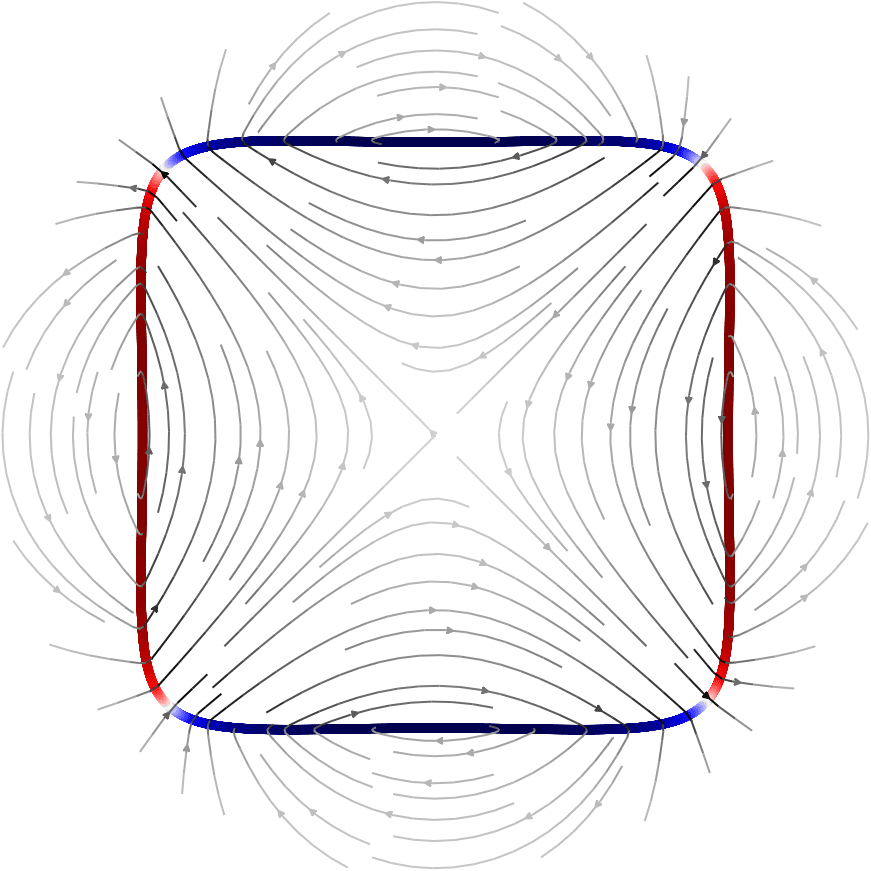}
		\caption*{\(B_2\) quadrupole}
		\label{square_B2}
	\end{minipage}
	\hfill
	\begin{minipage}{0.3\textwidth}
		\includegraphics[height=4.0cm]{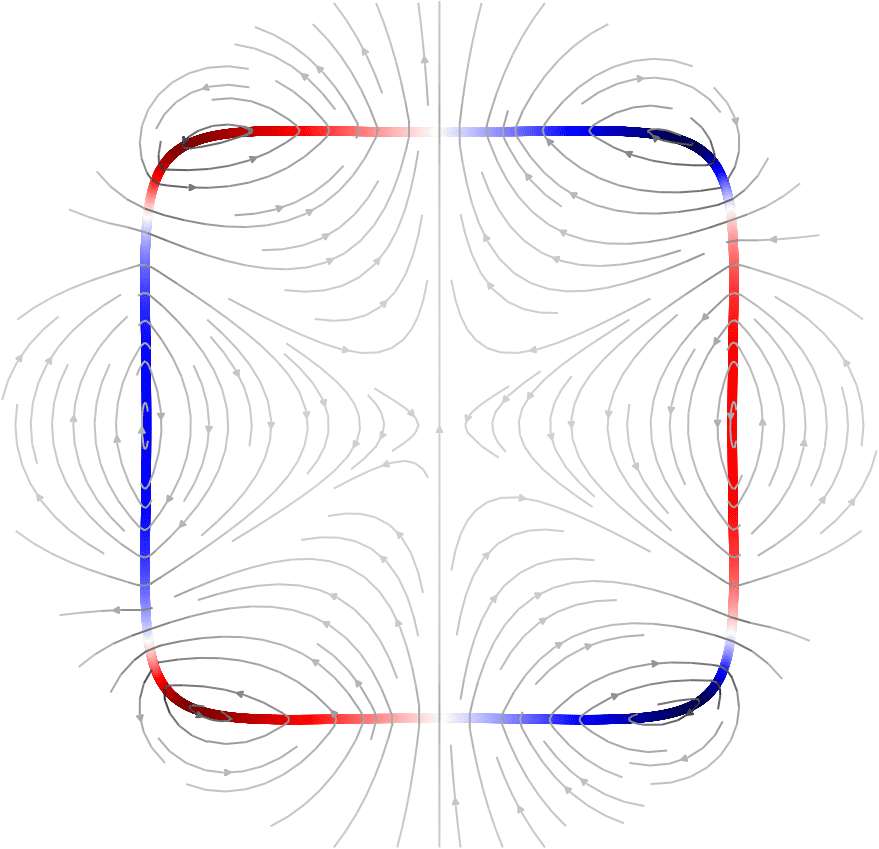}
		\caption*{\(B_3\) sextupole}
		\label{square_B3}
	\end{minipage}
	\caption{The current densities on a quasi-square shell to produce the first 3 circular field harmonics inside a quasi-square aperture.}
	\label{square_magnetic}
\end{figure*}

The transformation into a square form is expressed as:
\begin{equation}
	\zeta \left( z \right) =-\frac{z^3}{c^2}+\frac{c^2}{z}.
\end{equation}
Analogous to the triangular case, the square variant, which excludes the knotted segment for the domain that removes certain arcs, is represented by the following expressions:
\begin{equation}
	\begin{aligned}
		\frac{c}{3^{1/4}}<\rho <c,\Delta \theta & =\frac{1}{2}\arcsin \frac{c^4/\rho ^4-1}{2},\\
		\theta \in \left[ -\Delta \theta ,\Delta \theta \right] & \cup \left[ \frac{\pi}{2}-\Delta \theta ,\frac{\pi}{2}+\Delta \theta \right] \\
		\cup \left[ \pi -\Delta \theta ,\pi +\Delta \theta \right] & \cup \left[ \frac{3\pi}{2}-\Delta \theta ,\frac{3\pi}{2}+\Delta \theta \right].
	\end{aligned}
\end{equation}
The corresponding closed knotted curve in the \(\zeta\)-plane, mapped from the outer circle in the \(z\)-plane, is illustrated in \autoref{map_square_knotted}. 
Furthermore, the cut domain, which translates the constrained domain of definition of the \(z\)-plane into the entire domain of the \(\zeta\)-plane, is depicted in \autoref{map_square_shell}.
For a quasi-square shell, the current distribution, the resulting circular field harmonics inside, and the corresponding field outside are illustrated in \autoref{square_magnetic}.

When using conformal mapping to design magnets, careful selection of the former's radius, \(\rho_0\), in the \(z-\)plane is essential.
The radius \(\rho_0\) of the circular former's cross-section should not be too large, as this would lead to a quasi-polygonal former that no longer retains the properties of a convex polygon.
Conversely, \(\rho_0\)  should not be too small, as this would cause the quasi-polygonal former to approximate a circular shape, reducing the distinctiveness of the polygonal characteristics.

\begin{figure}[!htbp]
	\centering
	\includegraphics[width=7.5cm]{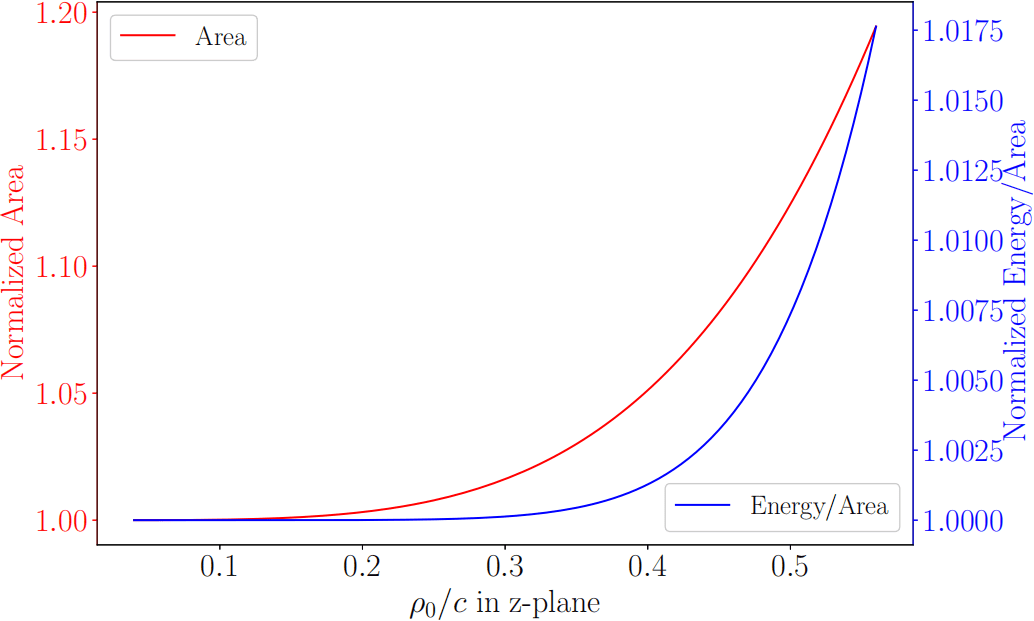}
	\caption{Stored energy and closed area as a function of \(\rho_0/c\) in \(z\)-plane. As \(\rho_0\) increases, the transformed shape changes from a circle to a quasi-square.}
	\label{Energy_fig}
\end{figure}

The magnetic stored energy per unit longitudinal length is given by:
\begin{equation}
	E=\frac{1}{2}\int{\left( \vec{B}\cdot \vec{H} \right) \text{d}{S}}=\frac{1}{2}\int{\mathbb{A}_z\mathbb{J}_z\text{d}{S}}=\frac{1}{2}\int{A_zJ_z\text{d}\theta}.
\end{equation}
We consider a square-like configuration corresponding to a contour at \(\rho_0\) in the \( z \)-plane. 
For various values of \(\rho_0\), we normalize the height and width of the transformed square shape to 1 and impose the field harmonics inside this shape as \(P \cos 2\Theta\) with the same magnetic gradient.
We calculate the areas of these square-like configurations and the corresponding magnetic field energy required for the current shell to excite, as depicted in \autoref{Energy_fig}.
As \(\rho_0\) increases, the shape gradually deforms from a circle to a square, resulting in an increase in the enclosed area. 
However, the stored energy per unit area required to sustain the magnetic field within this configuration remains nearly constant, with only a slight increase.
This observation implies that the square configuration can enclose a larger area while maintaining nearly the same efficiency in terms of current utilization.

\begin{figure}[!htbp]
	\centering
	\begin{minipage}{0.5\textwidth}
		\includegraphics[width=6.40cm]{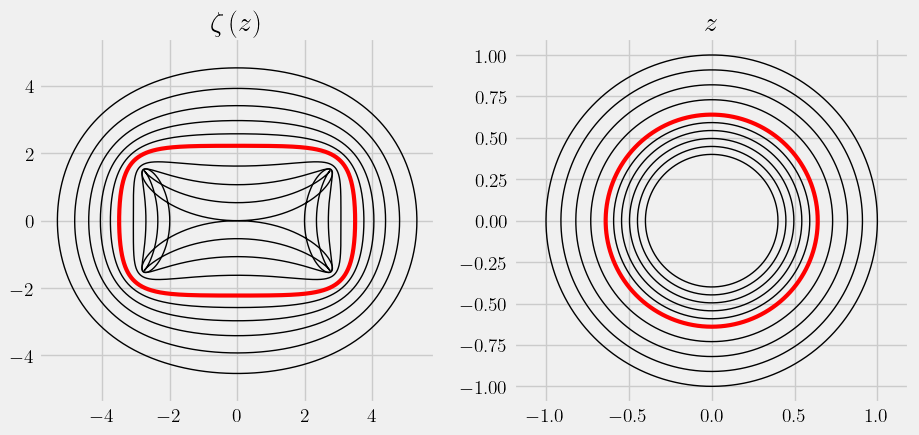}
		\caption*{conformal mapping}
		\label{rectangular_example}
	\end{minipage}
	\centering
	\begin{minipage}{0.5\textwidth}
		\includegraphics[width=4.8cm]{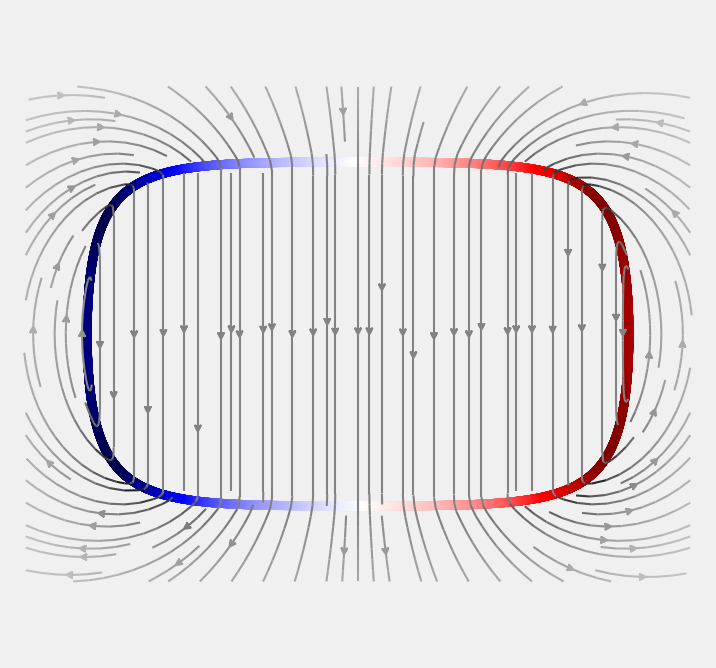}
		\caption*{\(B_1\) dipole}
		\label{Rec_dipole}
	\end{minipage}%
	\hfill
	\centering
	\begin{minipage}{0.5\textwidth}
		\includegraphics[width=4.8cm]{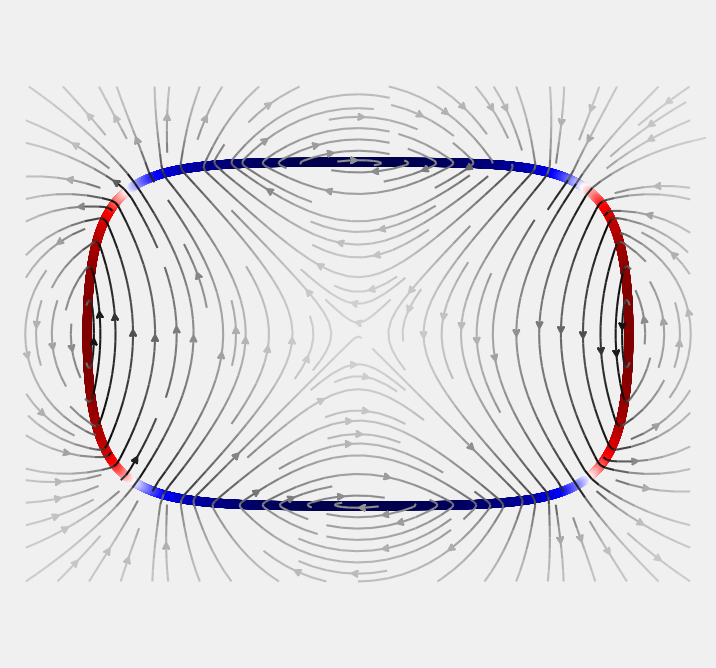}
		\caption*{\(B_2\) quadrupole}
		\label{Rec_quad}
	\end{minipage}
	\hfill
	\caption{Design of a rectangular former.}
	\label{rectangular_magnetic}
\end{figure}

In addition to the previously discussed shapes, apertures with other geometries can also be proposed, provided that the conformal mapping includes a \(1/z\) term and several polynomials in \(z\) with positive exponents.
In conformal mapping, power function transformations alter the shape of circles in the \( z \)-plane, while inversion transformations reverse the interior and exterior regions of a circle.
For instance, the transformation function for a quasi-rectangular shape is defined as:
\begin{equation}
	\zeta =-z^3+z+\frac{2}{z}.
\end{equation}
The conformal mapping generated by this function is illustrated in the first image of \autoref{rectangular_magnetic}.
Since the outer field outside the aperture is not of interest, we omit the process of constraining the domain of definition.
Using this transformation, the vector potentials for dipole and quadrupole fields inside quasi-rectangular apertures can be expressed as follows:
\begin{equation}
	\begin{aligned}
		P\cos \Theta &=\Re \left[ Pe^{i\Theta} \right]\\
		& =\Re \left[ -\rho ^3e^{i3\theta}+\rho e^{i\theta}+\frac{2}{\rho}e^{-i\theta} \right] \\
		&=-\rho ^3\cos 3\theta +\rho \cos \theta +\frac{2}{\rho}\cos \theta, \\
		P^2\cos 2\Theta &=\Re \left[ P^2e^{i2\Theta} \right]\\
		&=\Re \left[ \left( -\rho ^3e^{i3\theta}+\rho e^{i\theta}+\frac{2}{\rho}e^{-i\theta} \right) ^2 \right] \\
		&=\rho ^6\cos 6\theta -2\rho ^4\cos 4\theta -3\rho ^2\cos 2\theta\\
		&\quad + 4+\frac{4}{\rho ^2}\cos 2\theta.
	\end{aligned}
\end{equation}
As there is only one polynomial in \(\rho\) with a negative exponent, the corresponding current densities for the dipole and quadrupole fields are \(\mathbb{J}_{z} \sim \frac{\cos(\theta)}{|\zeta^\prime|}\) and  \(\mathbb{J}_{z} \sim \frac{\cos(2\theta)}{|\zeta^\prime|}\).
And the corresponding field can be seen in \autoref{rectangular_magnetic}.

\section{\label{App_CCT}Application to CCT magnets}
In \autoref{current_sheet}, we developed an analytical method to relate the desired circular field harmonics within a quasi-polygonal aperture to the current density distributed along a quasi-polygonal winding surface. 
In the following section, we will concentrate on methods to approximate this idealized current density using canted-cosine-theta (CCT) windings.

\begin{figure}[!htbp]
	\centering
	\includegraphics[width=5.6cm]{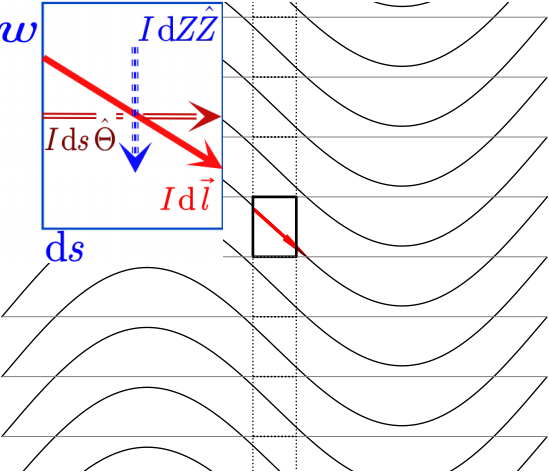}
	\caption{Spread the CCT (Canted Cosine Theta) winding on a cylindrical surface into a plane.}
	\label{CCT_illus}
\end{figure}

The key feature of CCT coils is the axial modulation of the winding path \( Z \), described by a periodic function superimposed with a constant pitch \(w\) per period, resulting in the helical trajectory characteristic of CCT windings, as shown in \autoref{CCT_illus}.
The trajectory of a generalized CCT coil is described by:
\begin{equation}
	\begin{aligned}
		&X = \Re\left[\zeta\left(z=\left(\rho_0, \theta\right)\right)\right], \\
		&Y = \Im\left[\zeta\left(z=\left(\rho_0, \theta\right)\right)\right], \\
		&Z = \frac{w}{2\pi} \theta + \frac{A_n}{n} \sin\left(n\theta\right), \\
		&\frac{\text{d}Z}{\text{d}s} = \frac{\text{d}Z/\text{d}\theta}{\text{d}s/\text{d}\theta} 
		= \frac{A_n \cos\left(n\theta\right) + \frac{w}{2\pi}}{|\zeta^\prime(z)|\rho_0}.
	\end{aligned}
	\label{CCT_path_formu}
\end{equation}
Here, \(\rho_0\) is the radius of a circular former in \(z-\)plane, \(\theta\) is the corresponding polar angle.
And \(\zeta\) represents the transformed coordinates in the \(\zeta-\)plane.
The transverse coordinates of winding  \(X, Y\) correspond to the real and imaginary parts of \(\zeta\), while \(s\) denotes the arc length projection in the transverse cross-section.
The modulation amplitude for the \(n-\)th harmonic is denoted by \(A_n\), and the function defining \(Z\) can incorporate any other sine or cosine terms, resulting in normal or skew magnetic fields.

To approximate the current distribution, a differential current element is decomposed into longitudinal and transverse components, as illustrated in \autoref{CCT_illus}.
By distributing the line current over the surrounding surface patch, a corresponding surface current density is obtained.
The longitudinal current density is given by \(\mathbb{J}_{z} = \frac{I}{w}\frac{\text{d}Z}{\text{d}s}\), while the azimuthal current density is \(\mathbb{J}_{\Theta} = \frac{I}{w}\), where \( I \) represents the total current and \( w \) denotes the winding pitch:
\begin{equation}
\mathbb{J}_{z} = \frac{I}{w}\frac{\text{d}Z}{\text{d}s},
\mathbb{J}_{\Theta} = \frac{I}{w}.
\label{CCT_current}
\end{equation}
Optimal winding pitch is essential to ensure the current distribution aligns with the analytical predictions, avoiding discontinuities from excessive pitch or manufacturing complexities from overly short pitch values.
In CCT magnet designs, where the tilt angle and current direction alternate between layers, the solenoidal field generated by the constant azimuthal current density \(\mathbb{J}_{\Theta} \) is effectively canceled within the aperture.
Simultaneously, the transverse field generated by the longitudinal current density \(\mathbb{J}_{z}\) is enhanced through constructive overlap between layers.

\begin{figure}[!htbp]
	\centering
	\includegraphics[width=6.4cm]{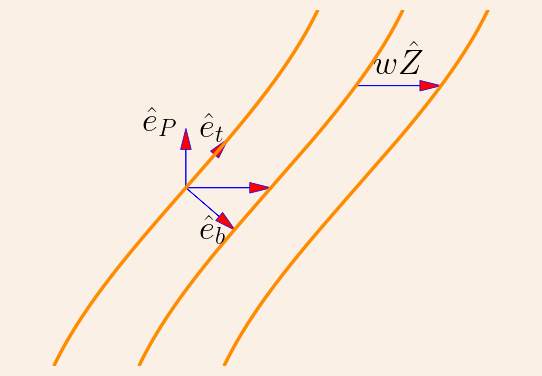}
	\caption{A local coordinate system to calculate the perpendicular distance \(\delta\) between adjacent CCT windings. Here, \(\hat{e}_t\), \(\hat{e}_P\), and \(\hat{e}_b\) correspond to the tangential, normal, and binormal directions, respectively, at any point on the wire.}
	\label{CCT_prove}
\end{figure}

A more rigorous method for averaging the winding current over the CCT surface can be found in~\cite{brouwer2015canted}. 
This method employs a local coordinate system to calculate the perpendicular distance between adjacent windings, visualized in \autoref{CCT_prove}. 
In a local coordinate system for a helical coil, the unit vectors 
\(\hat{e}_t\) , \(\hat{e}_P\), and \(\hat{e}_b\) correspond to the tangential, normal, and binormal directions, respectively, at any point on the wire. 
And \(\hat{\Theta}\) represents the unit tangential direction of the winding surface's transverse cross-section, and  \(\hat{Z}\) for longitudinal direction.
These unit vectors are typically defined based on the geometry of the coil's winding, and can be expressed as follows:
\begin{equation}
	\begin{aligned}
		\hat{e}_t &= \frac{\frac{\text{d}X}{\text{d}\theta}\hat{e}_x + \frac{\text{d}Y}{\text{d}\theta}\hat{e}_y + \frac{\text{d}Z}{\text{d}\theta}\hat{Z}}{\left| \frac{\text{d}X}{\text{d}\theta}\hat{e}_x + \frac{\text{d}Y}{\text{d}\theta}\hat{e}_y + \frac{\text{d}Z}{\text{d}\theta}\hat{Z} \right|} \\
		 &\sim \left( \rho_0 \left| \zeta' \right| \hat{\Theta} + \frac{\text{d}Z}{\text{d}\theta} \hat{Z} \right), \\
		\hat{e}_P &= \frac{\hat{Z} \times \hat{e}_t}{\left| \hat{Z} \times \hat{e}_t \right|} 
		\sim \hat{Z} \times \hat{\Theta}, \\
		\hat{e}_b  &= \hat{e}_t \times \hat{e}_P 
		\sim \left( \rho_0 \left| \zeta' \right| \hat{Z} - \frac{\text{d}Z}{\text{d}\theta} \hat{\Theta} \right),\\
		\hat{e}_b &= \frac{\rho_0 \left| \zeta' \right| \hat{Z} - \frac{\text{d}Z}{\text{d}\theta} \hat{\Theta}}{\left| \rho_0 \left| \zeta' \right| \hat{Z} - \frac{\text{d}Z}{\text{d}\theta} \hat{\Theta} \right|}.
	\end{aligned}
\end{equation}
Neglecting the wire's size and assuming tightly packed windings, the minimum spacing between adjacent turns can be expressed as:
\begin{equation}
	\delta = \left| w \hat{Z} \cdot \hat{e}_b \right| = w \left| \hat{Z} - \frac{\text{d}Z}{\text{d}\theta} \frac{\hat{\Theta}}{\rho_0 \left| \zeta' \right|} \right|^{-1}.
\end{equation}
When a current \(I\) is applied to the coil, the average surface current density is given by:
 \begin{equation}
 	\begin{aligned}
 	\vec{J} &= \frac{I}{\delta} \hat{e}_t = \frac{I}{w} \left| \hat{Z} - \frac{\text{d}Z}{\text{d}\theta} \frac{\hat{\Theta}}{\rho_0 \left| \zeta' \right|} \right| 
 	\times \frac{\hat{\Theta} + \frac{1}{\rho_0 \left| \zeta' \right|} \frac{\text{d}Z}{\text{d}\theta} \hat{Z}}{\left| \hat{\Theta} + \frac{1}{\rho_0 \left| \zeta' \right|} \frac{\text{d}Z}{\text{d}\theta} \hat{Z} \right|}\\
      & = \frac{I}{w} \hat{\Theta} + \frac{I}{w} \frac{1}{\rho_0 \left| \zeta' \right|} \frac{\text{d}Z}{\text{d}\theta} \hat{Z}.
    \end{aligned}
\end{equation}
The above analysis primarily considers the dominant current distribution within the CCT winding. 
For a more comprehensive and precise characterization of the magnetic field generated by the CCT winding, a detailed analysis is provided in \autoref{appen}.

\begin{figure}[!htbp]
	\centering
	\begin{minipage}{0.33\textwidth}
		\includegraphics[width=5.2cm]{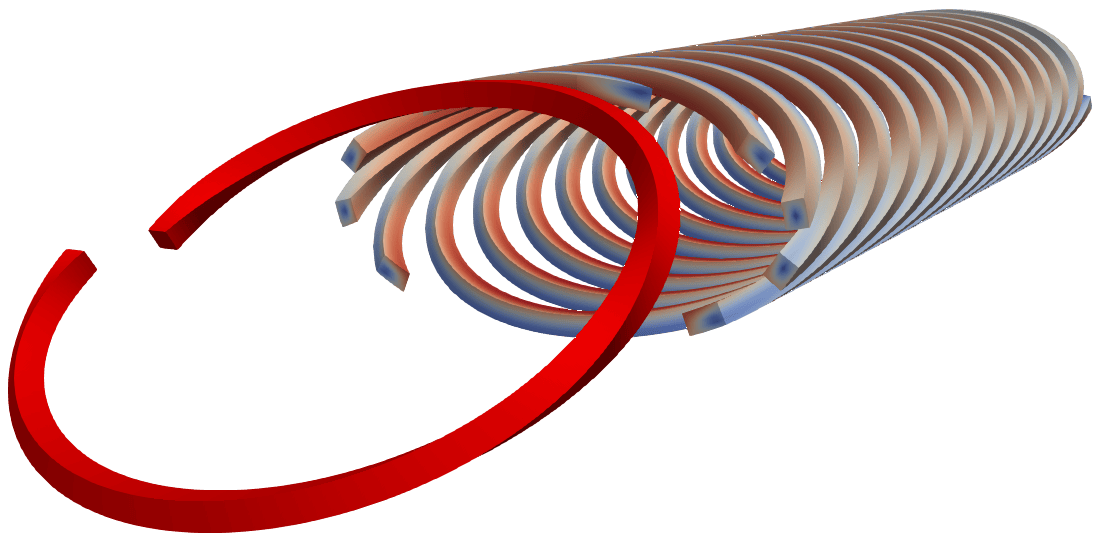}
		\caption*{Elliptic CCT winding for dipole}
		\label{Elliptic_dipole}
	\end{minipage}
	\hfill
	\begin{minipage}{0.33\textwidth}
		\includegraphics[width=5.2cm]{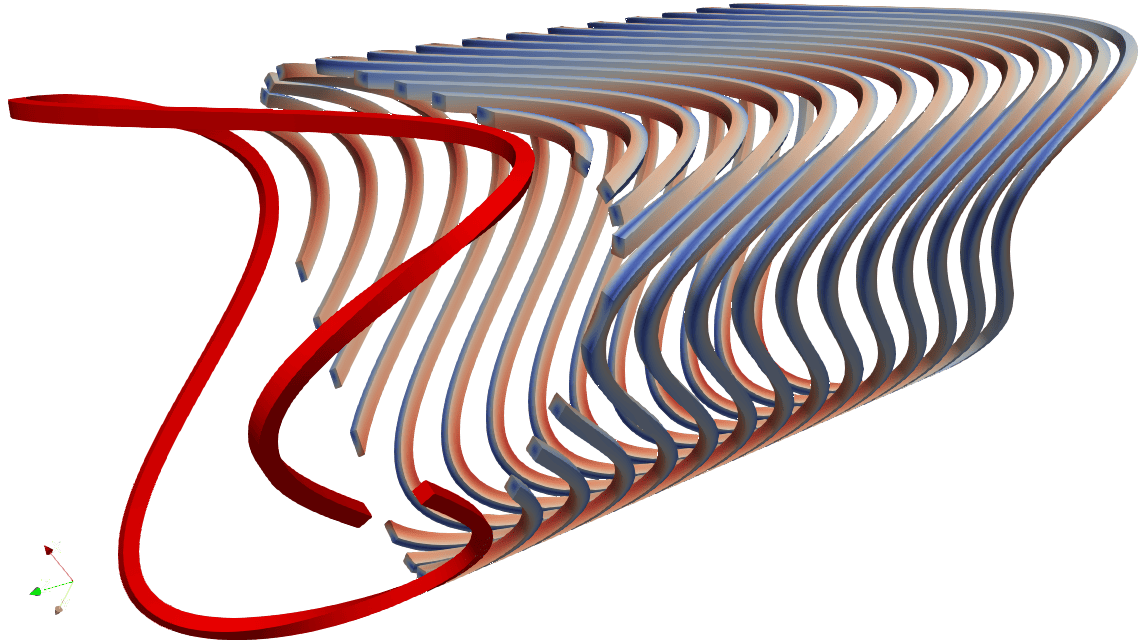}
		\caption*{Quasi-triangular CCT winding for sextupole}
		\label{Triangular_sextupole}
	\end{minipage}
	\hfill
	\begin{minipage}{0.33\textwidth}
		\includegraphics[width=5.2cm]{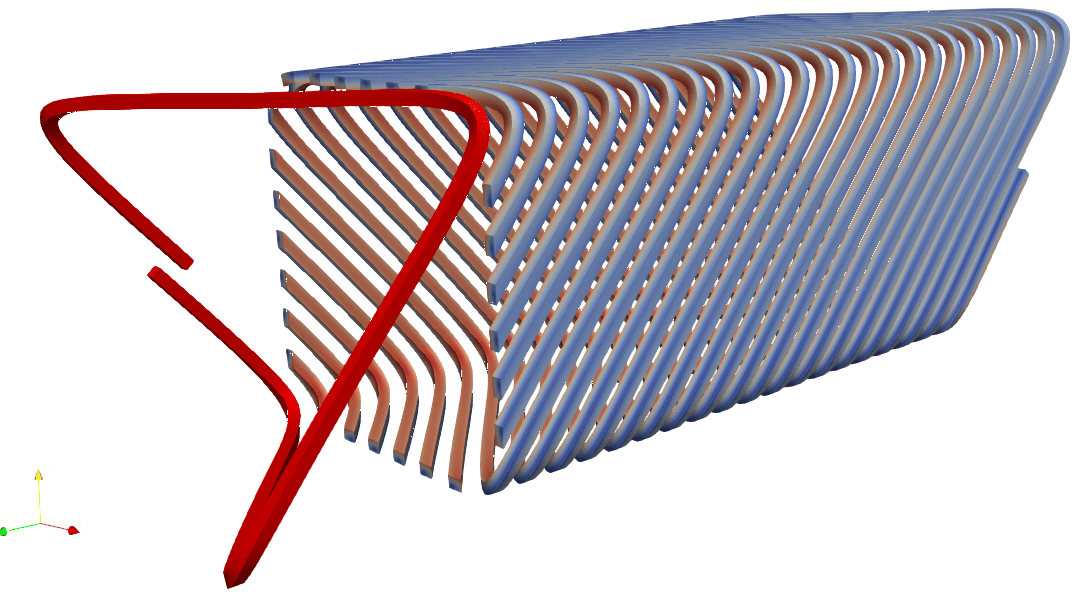}
		\caption*{Quasi-square CCT winding for quadrupole}
		\label{Square_quad}
	\end{minipage}
	\caption{Single layer quasi-polygonal CCT windings producing the circular field harmonics.}
	\label{cct_polygonal}
\end{figure}

The modeling of a CCT coil with quasi-polygonal magnets involves two key steps.
Firstly, to target the desired inner circular harmonics within a quasi-polygonal former, the corresponding current distribution \(\mathbb{J}_{z}\) is derived using conformal mapping techniques, as detailed in \autoref{current_sheet}.
Then the current distribution is  implemented through the CCT winding path provided in \autoref{CCT_path_formu}.
Certain complex magnetic field configurations may require multiple sine or cosine amplitude modulations to achieve the desired fields.
\autoref{cct_polygonal} presents an example of a quasi-polygonal coil former, designed to generate specific circular field harmonics within the magnetic structure. 
To check the accuracy of the results, we apply the Biot-Savart law to calculate the magnetic field generated by an idealized, infinitely long CCT coil with thin wires, confirming the presence and correctness of the field harmonics within the coil.
Given the finite longitudinal length of the actual CCT magnet and the non-negligible wire thickness, directly adopting winding parameters from analytical expressions is inadequate.
Consequently, in the engineering process, novel CCT designs with quasi-polygonal apertures necessitate detailed modeling to address geometric constraints and their influence on field uniformity. 
This optimization process is of comparable complexity to the methods already used in the design of high-current CCT layers with circular apertures~\cite{Ortwein_2021}.

\section{\label{Final}Conclusion}

This work presented a framework for the analytical design of superconducting magnets with quasi-polygonal bores. 
We began by deriving the relationship between quasi-polygonal and circular bores using conformal mapping. 
Next, we introduced a transformation of current distributions between these bores to achieve equivalent magnetic potentials.
Finally, this general approach was applied to the specific winding geometry of canted-cosine-theta (CCT) magnets, offering an analytical method for designing realistic coils for quasi-polygonal bore accelerator magnets.

\begin{acknowledgments}
	This work was supported by the National Natural Science Foundation of China (Grants No.12105005, No.12205007), and the National Grand Instrument Project (No.2019YFF01014403).
\end{acknowledgments}

\appendix
\section{\label{appen}Field distribution for CCT magnets}
\begin{figure}[!htbp]
	\centering
	\includegraphics[width=7.5cm]{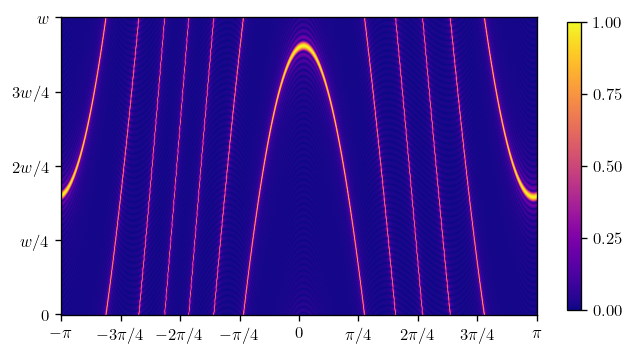}
	\caption{Approximation of the current distribution of a CCT dipole by truncating the delta function expansion to a finite number of terms \( |m|  < 24\).}
	\label{approximate_CCT}
\end{figure}

For an infinitely long, single-layer, straight CCT magnet, the winding trajectory can be parametrized by the wire's longitudinal variable \(Z\) and the trajectory's projection distance \(s\) within the transverse cross-section.
Due to the periodic nature of the winding along the longitudinal axis, \(Z\) is a periodic function of 
\(s\), with a period equal to the circumference of the cross-section.
Over each period, the winding advances a longitudinal distance of \(w\).
The current density distribution on the winding surface for a wire carry current \(I\) can then be expressed as:
\begin{equation}
	\vec{J}(s,z) =I\delta \left( z-Z \right) \hat{\Theta}+I\delta \left( z-Z \right) \frac{\text{d}Z}{\text{d}s}\hat{Z}.
\end{equation}

To reformulate this distribution, we expand the delta function in terms of a Fourier series:
\begin{equation}
	\begin{aligned}
		\delta \left( z-Z \right) &=\sum_m{\frac{e^{-im\frac{2\pi}{w}Z}}{w}e^{im\frac{2\pi}{w}z}}, \vec{J}=\sum_m{\vec{J}_m}, \\
		\vec{J}_m &=\frac{I}{w}e^{-im\frac{2\pi}{w}Z}\left( \hat{\Theta}+\frac{\text{d}Z}{\text{d}s}\hat{Z} \right) e^{im\frac{2\pi}{w}z}.
	\end{aligned}
\end{equation}
Here, \(\hat{\Theta}\) represents the unit tangential direction of the winding surface's transverse cross-section, while \(\hat{Z}\) denotes the unit vector in the longitudinal direction.
The current distribution for the CCT magnet can thus be approximated by retaining a finite number of terms in the delta function expansion, as illustrated in \autoref{approximate_CCT}.
Among these terms, \( \vec{J}_0 \) represents the commonly used approximation for the CCT current distribution, as mentioned in \autoref{CCT_current}. For the higher-order components, \( \vec{J}_m \) with \(m \neq 0\), the corresponding magnetic scalar potential \( \psi ^{\left( m \right)} \) satisfies the Helmholtz equation:
\begin{equation}
	\vec{J}_m\rightarrow \psi ^{\left( m \right)}e^{im\frac{2\pi}{w}z},		\nabla^2  \psi ^{\left( m \right)} =\frac{4\pi ^2m^2}{w^2}  \psi ^{\left( m \right)}.
\end{equation}

In the case of a CCT magnet with circular former, this Helmholtz equation can be solved in polar coordinates, yielding solutions involving the modified Bessel functions of the first \(  I_{|n|}\left(  \left| \frac{2\pi m}{w} \right| \rho \right)  \) and
second kind \(K_{|n|}\left(  \left| \frac{2\pi m}{w} \right| \rho \right)\) to describe the magnetic field both inside and outside the winding surface~\cite{tuprints11687}.
The corresponding expressions are:
\begin{equation}
	\begin{aligned}
		\psi ^{\left( m \right)}&=\sum_n{\psi ^{\left( n,m \right)}e^{in\theta}},\\
		\frac{1}{\rho}\partial _{\rho}\left( \rho \partial _{\rho}\psi ^{\left( n,m \right)} \right) &=\left( \frac{n^2}{\rho ^2}+\frac{4\pi ^2m^2}{w^2} \right) \psi ^{\left( n,m \right)}.
	\end{aligned}
\end{equation}

\begin{equation}
	\left\{
	\begin{aligned}
		\text{inner:\,\,}\psi ^{\left( n,m \right)} &=\phi _{\text{in}}^{\left( n,m \right)}I_{|n|}\left(\left| \frac{2\pi m}{w} \right|
		\rho \right) ,\\
		\text{outer:\,\,}\psi ^{\left( n,m \right)} &=\phi _{\text{out}}^{\left( n,m \right)}K_{|n|}\left( \left| \frac{2\pi m}{w}  \right|\rho\right) .
	\end{aligned}
	\right.
\end{equation}

The coefficients \(\phi _{\text{in}}^{\left( n,m \right)}\) and 
\(\phi _{\text{out}}^{\left( n,m \right)}\) can be determined by applying  boundary conditions for the magnetic field at the winding surface, taking into account the current distribution along the winding \(\vec{J}_m\).
But for a CCT magnet with a quasi-polygonal former, the solution to the Helmholtz equation lacks a concise analytical form.

\pagebreak
\newpage
\pagebreak
\newpage
\newpage
\nocite{*}


%

\end{document}